\documentclass[useAMS,usenatbib]{mn2e}
\usepackage{graphicx}
\usepackage{amssymb}
\usepackage{latexsym}
\usepackage{keyval}
\usepackage{amsmath}

\usepackage{natbib}
\usepackage{cite}

\newcommand{\degree}{^{\circ}}

\title[Young stellar cluster in the outer Galaxy]{Not a galaxy: IRAS~04186+5143, a new young stellar cluster in the outer Galaxy}
\author[J. L. Yun, D. Elia, A. A. Djupvik, J. M. Torrelles, and S. Molinari]{ J. L. Yun$^{1,2}$\thanks{E-mail: yun@oal.ul.pt}, Davide Elia$^{3}$\thanks{E-mail: davide.elia@ifsi-roma.inaf.it}, A. A. Djupvik$^{4}$\thanks{E-mail: amanda@not.iac.es}, J. M. Torrelles$^{5}$\thanks{E-mail: torrelles@ieec.cat \newline {(The ICC (UB) is a CSIC-Associated Unit through the ICE)}},  S. Molinari$^{3}$\thanks{E-mail: sergio.molinari@ifsi-roma.inaf.it}\\
$^{1}$Instituto de Astrof\'{\i}sica e Ci\^encias do Espa\c{c}o -- Universidade de Lisboa, Observat\'orio Astron\'omico de Lisboa, Tapada da Ajuda, \\ 1349-018 Lisboa, Portugal\\
$^{2}$Departament d'Astronomia i Meteorologia, Institut de Ci\`encias del Cosmos, Universitat de Barcelona (IEEC-UB), Mart\'{\i} i Franqu\`es 1, \\ E-08028 Barcelona, Spain\\
$^{3}$INAF - Istituto di Astrofisica e Planetologia Spaziali, via Fosso del Cavaliere 100, 00133, Rome, Italy\\
$^{4}$Nordic Optical Telescope, Nordic Optical Telescope, Rambla Jos\'{e} Ana Fern\'{a}ndez P\'{e}rez, 7,
              ES-38711 Bre\~{n}a Baja, Spain \\
$^{5}$Institut de Ci\`encies de l'Espai (CSIC-IEEC) and Institut de Ci\`encies del Cosmos (UB-IEEC),  Mart\'{\i} i Franqu\`es 1, E-08028 Barcelona, \\ Spain \\
}

\begin{document}

\date{ Accepted 2015 June 24. Received 2015 June 23; in original form 2015 April 29}

\pagerange{\pageref{firstpage}--\pageref{lastpage}} \pubyear{2015}

\maketitle

\label{firstpage}

\begin{abstract}
We report the discovery of a new young stellar cluster in the outer Galaxy located at the position of an IRAS PSC source that has been previously mis-identified as an external galaxy. The cluster is seen in our near-infrared imaging towards IRAS~04186+5143 and in archive Spitzer images confirming the young stellar nature of the sources detected. There is also evidence of sub-clustering seen in the spatial distributions of young stars and of gas and dust.

Near- and mid-infrared photometry indicates that the stars exhibit colours compatible with reddening by interstellar and circumstellar dust and are likely to be low- and intermediate-mass YSOs with a large proportion of Class~I YSOs. 

Ammonia and CO lines were detected, with the CO emission well centred near the position of the richest part of the cluster. The velocity of the CO and NH$_3$ lines indicates that the gas is Galactic and located at a distance of about 5.5 kpc, in the outer Galaxy. 

{\it Herschel} data of this region characterise the dust environment of this molecular cloud core where the young cluster is embedded. We derive masses, luminosities and temperatures of the molecular clumps where the young stars reside and discuss their evolutionary stages. 
\end{abstract}

\begin{keywords}
stars: formation -- infrared: stars -- submillimetre: ISM -- ISM: clouds -- ISM: Individual objects: IRAS~04186+5143  -- ISM: dust, extinction.
\end{keywords}

\section{Introduction}

It has been well established that star formation occurs across the Galactic disc and at different Galactocentric distances. Both in the inner and in the outer Galaxy, young stellar clusters still partly embedded in the dense gas and dust in molecular clouds have been found \citep[e.g.][]{tapia91, strom93, mccaughrean94, horner97, luhman98, santos00}. They represent current active star formation sites. The star formation activity seen throughout the Galactic disc is possible due to the relatively large amounts of dust that shield young forming stars from the heating of the external interstellar radiation field. The dust produces high values of extinction resulting in lines of sight across the Galactic disc that are highly opaque in the optical wavelengths. 

Among other effects, the large values of dust extinction along the Galactic disc make difficult the task of achieving a complete census of the Milky Way neighbour satellite galaxies. Even with our best instruments, we may not have found and catalogued correctly all the stellar systems components of the Milky Way and its neighbour galaxies. Conversely, Galactic stellar systems can be wrongly classified as extragalactic neighbours. As an example of this fact, \citet{martin04} have claimed the discovery of remains of a satellite dwarf galaxy, a claim that was subsquently challenged \citep{momany06}. More recently, the search for dwarf satellite galaxies, both of 
the Milky Way and of Andromeda continues \citep[e.g.][]{conn12,sesar14}.

The detection and characterisation of star formation sites in an early stage have strong implications on the structure and evolution of the Galaxy. However, the study and census of star formation sites in the outer Galaxy, and specially at large distances, has received less attention and coverage when compared to the inner Galaxy and the solar neighborhood.

IRAS~04186+5143 is an IRAS PSC source in the outer Galaxy that appears classified in the SIMBAD data base as ``2MASX J04223304+5150346 -- Galaxy''. This means that it is listed in 2MASX (the 2Micron All-Sky Survey extended source catalogue) as being an extragalactic source. It is also an extended submm and far-infrared source, having been detected in the submm continuum and listed in the SCUBA legacy catalogues \citep{difrancesco03}, and also detected by the {\it Herschel} satellite \citep{ragan12}.  In addition, mid-infrared spectral features have been seen towards this region (ISOSS J04225+5150 East) using  {\it Spitzer} \citep{pitann11}, and \citet{birkmann07} has derived a kinematic distance of 5.5 kpc. Furthermore, \citet{sunada07} found no water maser emission in their survey. All the authors above clearly refer to this source as a Galactic object.

As part of our study of young embedded clusters in the outer Galaxy \citep[e.g.][]{yun09,palmeirim10},
we have conducted observations (near-infrared $JHK_S$ imaging, and millimetre CO line) towards IRAS~04186+5143. These observations revealed the 
presence of a young stellar population embedded in a molecular cloud core. 
We report here our near-infrared discovery of a young stellar cluster seen towards IRAS~04186+5143, and exhibiting evidence of sub-clustering. In addition, we use new CO data, as well as {\it Herschel} observations, archive {\it Spitzer}, and archive ammonia VLA data to characterise the molecular environment and the young stellar population. 
Section~2 describes the observations and data reduction. In Section~3, we present and discuss the results. A summary is given in Section~4.

\section[]{Observations and data reduction}

\subsection{Near-infrared observations}

Near-infrared ($J$, $H$, and $K_S$) images were obtained on September 8th 2009 using the Nordic
Optical Telescope near-IR Camera and Spectrograph (NOTCam). The detector was the 1024 $\times$
1024 $\times$ 18 micron Hawaii science grade array (SWIR3). The wide-field camera (0.234$''$/pix)
was used, and the observations were performed using a ramp-sampling readout mode. Every
sky position was integrated for 36 ($K$) or 48 ($J$ and $H$) seconds, reading out the array
every 6 ($K$) or 8 ($J$ and $H$) seconds, and using the linear regression result of the 6
readouts. The raw images were corrected for non-linearity using a pixel-by-pixel
correction model available for NOTCam. Differential twilight flats were used for
flat-fielding. All images were bad-pixel corrected, flat-fielded, sky-subtracted,
distortion corrected (using a model of the WF-camera distortion), shifted, and combined
to one deep image per filter. The total integration time in the final images is 648 ($K$)
and 816 ($J$ and $H$) seconds.

Point sources were extracted using {\tt daofind} with a detection threshold of 
5$\sigma$. The images were then examined for false detections and a few sources were eliminated by hand. Aperture 
photometry was made with a small aperture (radius = 3 pix, which is about
the measured full-width-at-half-maximum of the point spread function) and aperture corrections, 
found from 20 bright and isolated stars in each image, were used to correct 
for the flux lost in the wings of the PSF. The error in  determining 
the aperture correction was $<$ 0.02 mag in all cases. These errors were added to the value MERR, which is created by the IRAF task {\em phot}. 
A total of 848, 984, and 786 sources were found to have fluxes in $J$, 
$H$, and $K_S$, respectively, and errors $\sigma_{Ks} <$ 0.25 mag.

We used the 2MASS All-Sky Release Point Source Catalogue \citep{cutri03,skrutskie06} to calibrate our observations.
The $JHK_S$ zeropoints were determined using 2MASS stars brighter than $K_S$ = 14.5 mag.
The standard deviations of the offsets between NOTCam and 2MASS photometry
are 0.04, 0.06, and 0.05 mag, in $J$, $H$, and $K_S$, respectively. We estimate the completeness limit of the observations to be roughly 19.0, 18.5, and 18.0 magnitudes in $J$, $H$, and $K_S$, respectively.

\subsection{Millimetre line observations}
The region around the position of the IRAS source was mapped using the single-dish OSO 20-m radiotelescope (Onsala, Sweden) in 2009 April. Three maps were obtained in the rotational lines of $^{12}$CO(1-0), $^{13}$CO(1-0), and CS(2-1) at 115.271, 110.201, and 97.981~GHz, respectively. Since the telescope half-power beam width (HPBW) is 33\arcsec at 115 GHz, we decided to obtain the maps with a grid spacing of 30\arcsec, centered on the IRAS coordinates and composed by $5 \times 5$ pointings for the two CO lines, and by $3\times3$ pointings for the CS(2-1) line. The typical integration time was 120, 240, and 300~s for  $^{12}$CO(1-0), $^{13}$CO(1-0), and CS(2-1), respectively.

A high resolution 1600-channel acousto-optical spectrometer was used as a back end, with a total bandwith of 40 MHz and a channel width of 25 KHz that, at the observed frequencies, corresponds to 0.065, 0.068, and 0.076 km~s$^{-1}$, respectively. The spectra were generally taken in dual beam switching mode, except for the $3 \times 3$ innermost portion of the $^{13}$CO(1-0) map, observed in frequency switching mode. The antenna temperature was calibrated with the standard chopper wheel method. Pointing was checked regularly towards known circumstellar SiO masers; pointing accuracy was estimated to be 3\arcsec rms in azimuth and elevation.

The data reduction consisted of a typical pipeline for mm spectra: first, a folding operation was applied only to frequency-switched spectra; then, the baseline has been fitted by a third-order polynomial, and subtracted from the spectra (the resulting rms noise per channel is 0.83, 0.35, and 0.09~K for $^{12}$CO(1-0), $^{13}$CO(1-0), and CS(2-1), respectively); finally, in all spectra the antenna temperature $T_{\mathrm A}$ was translated in main beam temperature $T_{\mathrm {MB}}$ dividing by the telescope main beam efficiency factor $\eta_{\mathrm {MB}}$. This parameter is generally quoted as a constant of the telescope, but instead it can vary with the elevation of the source; since this variation is evaluated and provided by the OSO 20m telescope system at each pointing $\mathrm (i,j)$, we chose to divide each spectrum by its peculiar $\eta_\mathrm {MB}(i,j)$ value.

\subsection{VLA ammonia observations}

Simultaneous observations of the NH$_3$(1,1) and NH$_3$ (2,2) lines (rest frequencies 23.694495 GHz and 23.722633 GHz, respectively) were carried out with the Very Large Array (VLA)  of the National Radio Astronomy Observatory (NRAO)\footnote{The NRAO is a facility
of the National Science Foundation operated under  cooperative agreement by Associated
Universities, Inc.} in the D configuration during 2003 April 19 (project AK562; NRAO public archive data). A bandwidth of 3.1~MHz with 63 spectral channels of 48.8~kHz width ($\sim$0.62~km~s$^{-1}$ at $\lambda$ = 1.3~cm) was selected for each ammonia line. The center channel velocity was set at V$_{LSR}$ = $-$43.7~km~s$^{-1}$, covering a total velocity range $-$62.9~km~s$^{-1}$ $\lesssim$ V$_{LSR}$ $\lesssim$ $-$24.5~km~s$^{-1}$. The absolute coordinates of the phase center were $\alpha$(J2000) = 04$^h$22$^m$34.358$^s$, $\delta$(J2000) = 51$^{\circ}$50$'$51.0$''$, which is $\sim$20$''$ northeast from the nominal position of IRAS~04186+5143. The observing on-source time was $\sim$3~hours. 0542+498 was used as flux calibrator, assuming a flux density of 1.78~Jy at 1.3~cm.
The phase calibrators were 0359+509 and 4C50.11, with bootstrapped flux densities 9.04$\pm$0.02~Jy and 8.46$\pm$0.05 Jy at 1.3~cm, respectively. Calibration and imaging was made using the Astronomical Image Processing System (AIPS) software of NRAO. The resulting synthesised beam size was $\sim 3.3'' \times 2.9''$ (p.a. = 78$^{\circ}$) with the uv data naturally weighted. An rms per spectral channel of $\sim$1.8~mJy~beam$^{-1}$ was obtained in the images.
We estimate that the absolute positions are accurate to $\sim$0.5$''$.

\subsection{Spitzer observations}
We searched the \textit{Spitzer} data archive, and found observations of the IRAS~04186+5143 region in all the four bands (3.6, 4.5, 5.8, and 8~$\mu$m) of the InfraRed Array Camera \citep[IRAC,][]{faz04}, and at 24~$\mu$m of Multiband Imaging Photometer for \textit{Spitzer}, being part of the program PID 20444, executed on 2005 September 20. For IRAC observations, we used the basic calibrated data (BCD) images produced by the S18.7.0 pipeline of the \textit{Spitzer} Science Center: 96 dithered frames with a 10.4~s exposure, and 32 with 0.4~s are available. After having removed the residual muxbleed artifacts from the single frames \citep{hor04}, we combined into mosaics
using the MOPEX software \citep{mak05a} to obtain
two mosaics for each band, corresponding
to the long- and to the short-exposure time.
The final maps have a size of $\sim 5\arcmin.5 \times
5\arcmin.5$, and a scale of $0\arcsec.6/$pixel.

Point source detection and photometry extraction was
performed with MOPEX as well \citep{mak05b}, independently
at each band. When a source is present both in the long-
and in the short-exposure image, the photometry taken from
the latter is considered as more reliable. After band merging
(based on simple spatial association) a four-band catalog of
1020 entries (having at least a detection in one of the bands)
has been obtained. In particular, sources with detections at
the four bands are 215, whereas sources detected in bands 2, 3
and 4 are 221.

\subsection{Herschel observations}

IRAS 04186+5143 was observed in the far-IR within the \emph{Herschel Infrared Galactic Plane Survey} 
\citep[HI-GAL,][]{mol10}, a Herschel open-time key project which mapped the Galactic plane with 
the Photodetector Array Camera and Spectrometer \citep[PACS, 70 and 160 $\mu$m;][]{pog10} and the 
Spectral and Photometric Imaging Receiver \citep[SPIRE, 250, 350 and 500 $\mu$m;][]{gri10} instruments on board the
Herschel satellite \citep{pil10}. The HI-GAL observations are arranged in \emph{tiles} of 
$\sim 2.3\degree \times 2.3\degree$ taken at each of the five wavelengths. IRAS 04186+5143 
can be found in the HI-GAL field centred at $[\ell,b]=[152\degree,+1\degree]$ and identified as Field$\_151\_0$ 
in the Herschel Science Archive, observed by Herschel on February 13th 2012 in PACS+SPIRE parallel mode at 
a scan speed of 60 arcsec s$^{-1}$. The data were reduced using the UNIMAP pipeline, a map maker developed within the Hi-GAL project \citep{pia15}. The maps have pixel sizes 3.2, 4.5, 6, 8 and 11.5 arcsec at 70, 160,
250, 350, 500 $\mu$m, respectively. As in \citet{eli13}, the astrometry of the maps was checked by comparing the positions 
of several isolated compact sources appearing in both the 70~$\mu$m map and in the WISE
survey \citep{wri10} at 22 $\mu$m. Finally, a zero-level offset, obtained by comparing the Herschel data with Planck and IRAS data, following \citet{ber10}, was evaluated 
and added to Herschel maps at each band.

Compact source extraction and photometry have been performed using the Curvature Threshold Extractor package \citep[CuTEx,][]{mol11}, adopting the same prescriptions and settings used for the general Hi-GAL compact source catalog (Molinari et al. 2014, in prep.). The subsequent band merging procedure has been carried out based on simple spatial association criteria \citep[see, e.g.][]{eli10}.

\begin{figure}
   \centering
   \includegraphics[width=8cm]{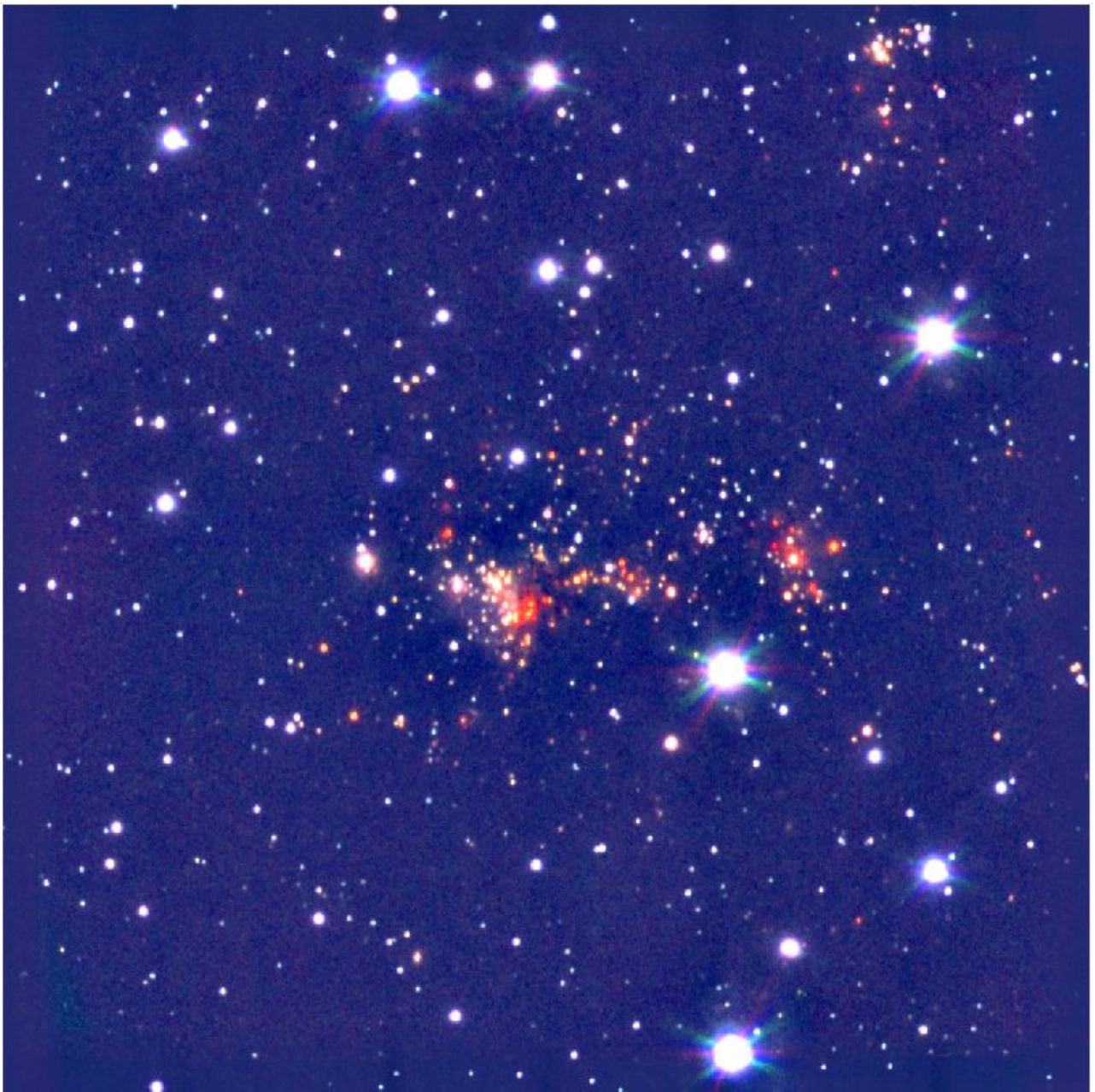}
   \caption{$J$ (blue), $H$ (green), and $K_S$ (red) colour composite image towards IRAS~04186+5143 covering $4' \times 4'$. North is up and East to the left. Note the concentration of red stars towards the centre of the image and in the northwest corner. The central concentration itself exhibits two groups (sub-clusters) of red stars. }
   \label{JHK}
\end{figure}

\begin{figure}
   \centering
   \includegraphics[width=9cm]{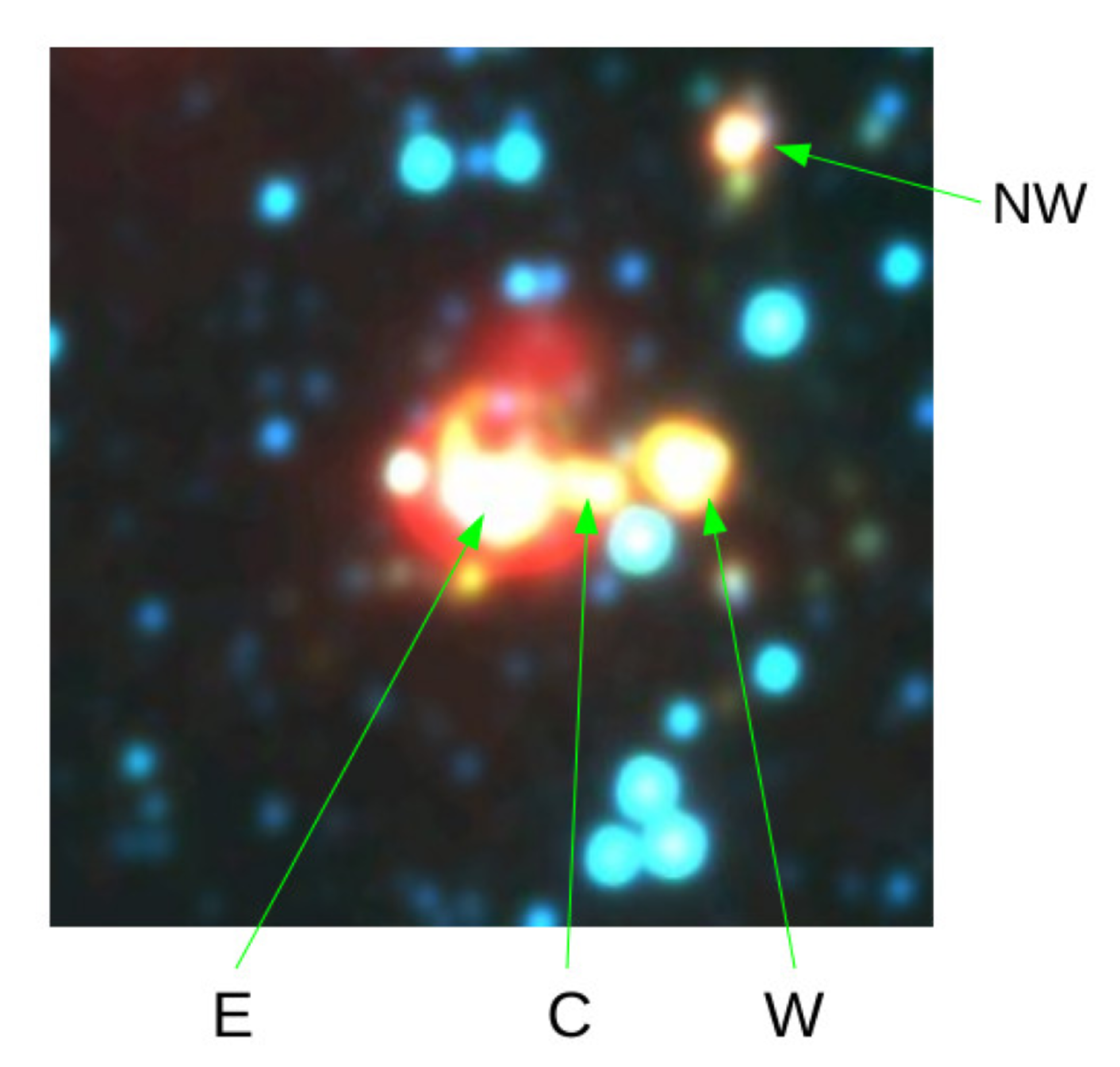}
   \caption{WISE colour composite image centred at IRAS~04186+5143 covering $5' \times 5'$. North is up and East  to the left. Sub-clusters ``E'', ``W'', and ``NW'', mentioned in the text, are marked. ``E'', ``W'', ``C'', and ``NW'' represent also the clumps detected in Herschel's images (see below).  }
   \label{WISE}
\end{figure}

\begin{figure}
   \centering
   \includegraphics[width=9cm]{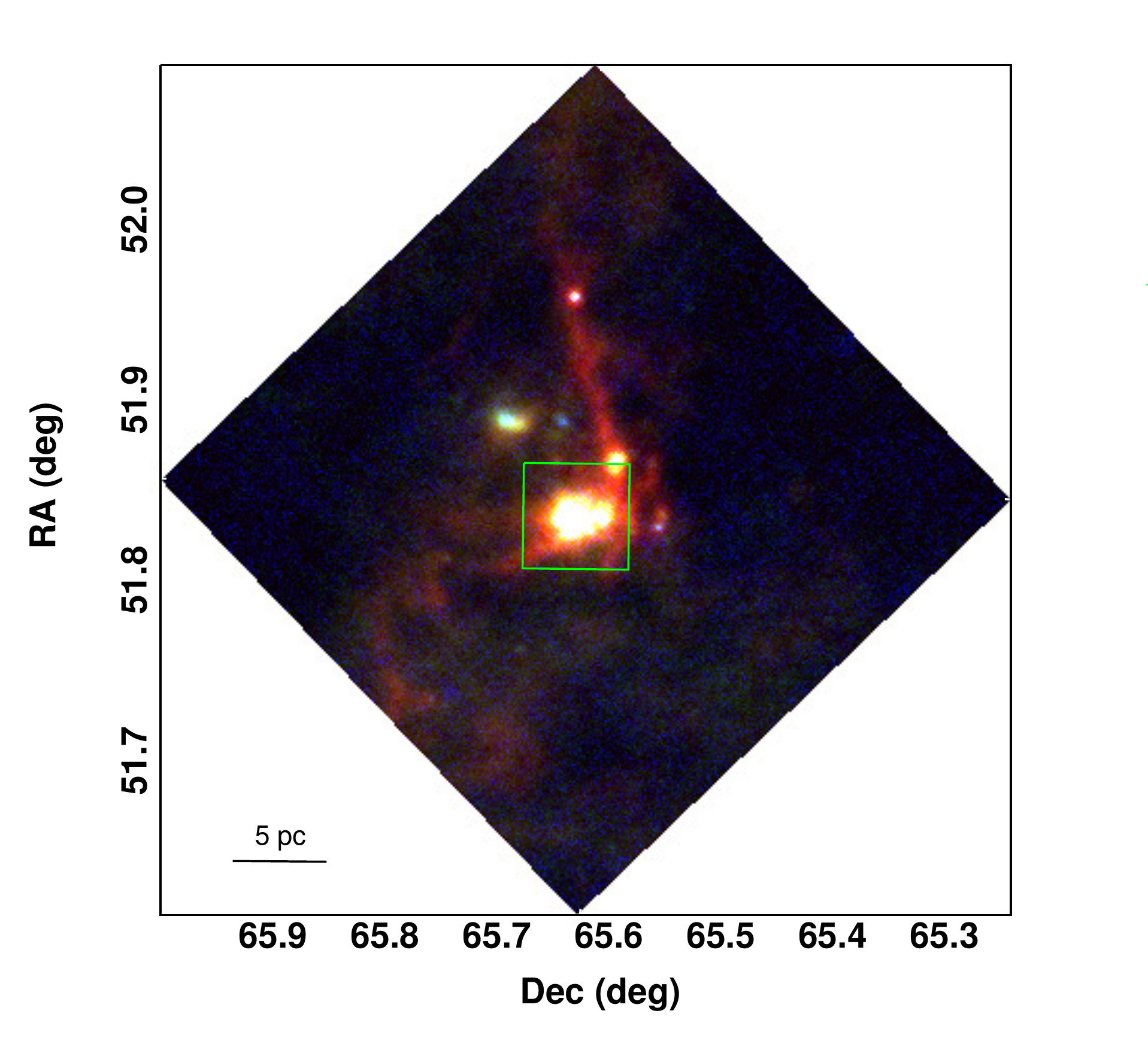}
   \caption{Large-scale Hi-GAL three-colour composite image (\textit{blue}, 70~$\mu$m; \textit{green}, 160~$\mu$m; \textit{red}, 350~$\mu$m) of the region around IRAS~04186+5143. The green box encompasses the area observed in the near-infrared. The ruler at the bottom-left corner corresponds to 5~pc (at a distance of 5500~pc).}
   \label{hls}
\end{figure}

\section{Results and discussion}

\subsection{The infrared morphology}

Figure~\ref{JHK} presents the NOTCam $JHK_S$ near-infrared colour composite image obtained towards IRAS~04186+5143. A higher concentration of ``red'' stars (much brighter in the $K_S$-band than in the $J$ or $H$-bands) is seen close to the centre of the image. The ability to resolve most stars in this concentration (with possible exceptions at the most crowded region) argues in favour of these sources being Galactic. In addition, the location of this concentration of stars coincides with the position where the molecular gas, traced by CO, peaks (see below), marking the presence of a molecular clump. This good spatial coincidence of red stars and molecular gas strongly argues in favour of their association. Thus, this image reveals a young stellar cluster still embedded in a dense cloud core located in the outer Galaxy.

Under a closer look, Figure~\ref{JHK} hints at the presence of sub-clustering. In fact, the red stars appear to cluster around the centre of the frame, but also around a more western point. Interestingly, as we show below, the column density of the molecular cloud core seems to have a secondary peak west of the centre, coincident with the location of the western red stars. 
Furthermore, a third smaller group of red stars is seen towards the northwest corner if the image. These red sources are also seen in the WISE \citep{wri10} archive data base colour composite image shown in Figure~\ref{WISE}.

Moving to the far-infrared, Figure~\ref{hls} shows a {\it Herschel} RGB (70-160-350)~$\mu$m composite map of this region at a larger scale.  The green square indicates the region observed by the NOT. IRAS~04186+5143 appears embedded in its environment, with fainter filaments connecting bright clumps and joining at the location where star formation is most active \citep[cf.][]{sch12}, a morphology typical of a Galactic star forming region.

\subsection{Molecular gas morphology and kinematics}

\begin{figure*}
   \centering
   \includegraphics[width=7.6cm]{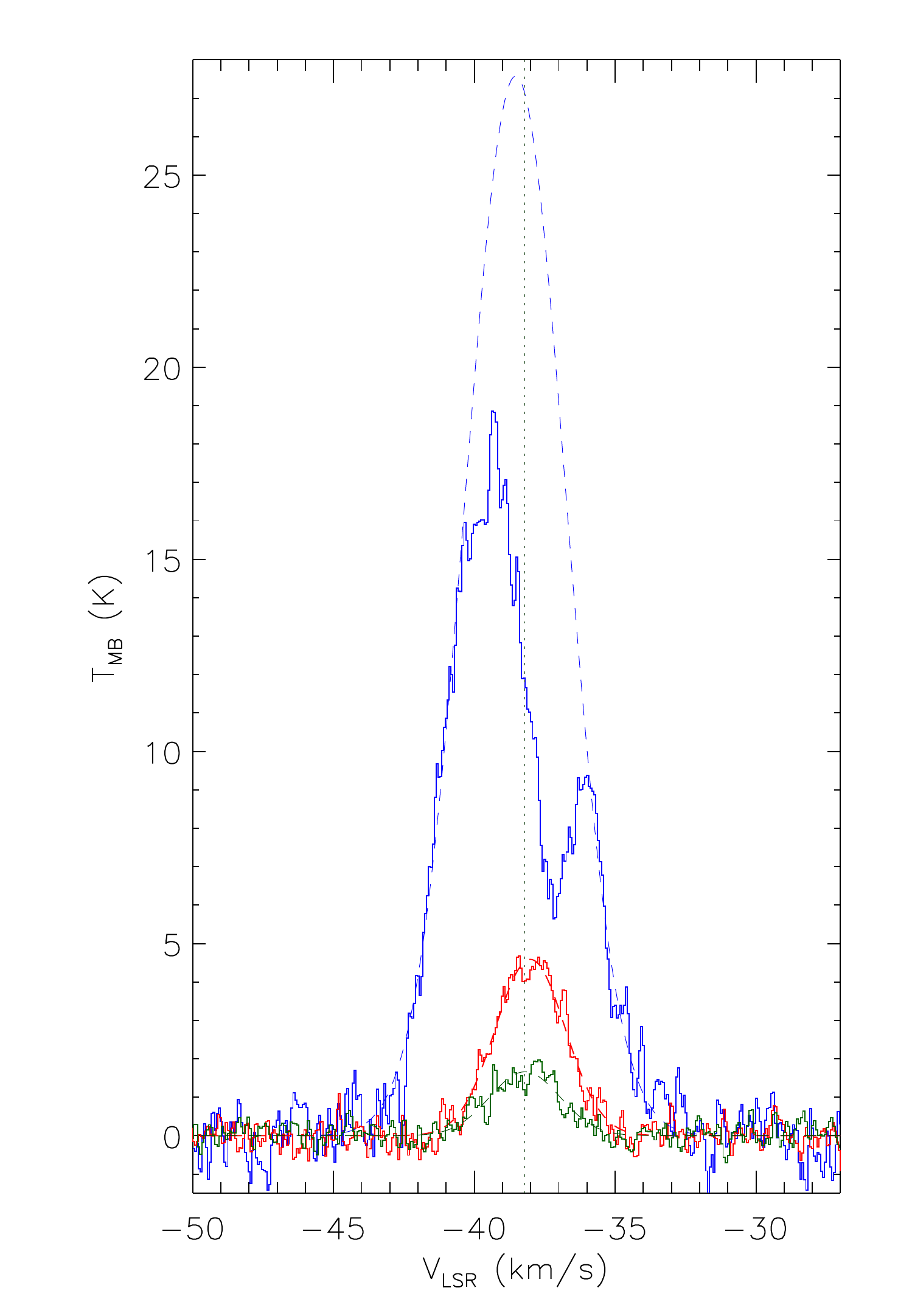}
   \caption{Spectra of the three observed lines toward the centre position of the CO map, corresponding to the location of the IRAS source. Blue: $^{12}$CO(1-0); red: $^{13}$CO(1-0); green: CS(2-1). The dashed lines represent the Gaussian fits of the observed lines (see text). The vertical grey dotted line indicates the peak position of the Gaussian fit of the CS(2-1) line.
   }
   \label{pos00}
%\end{figure}

%\begin{figure*}
%\begin{figure}
   \centering
   \includegraphics[width=10cm]{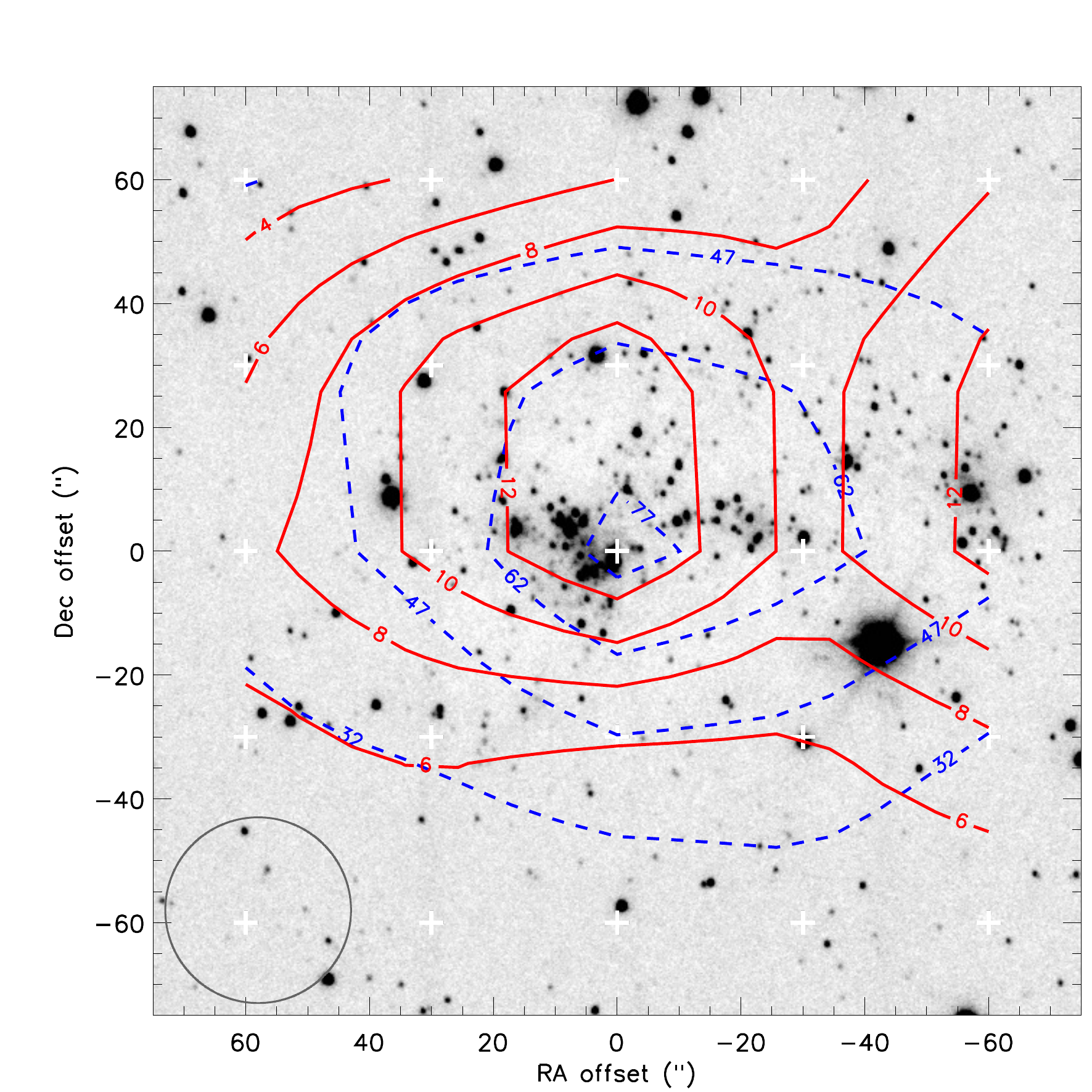}
   \caption{Contour maps (smoothed) of the $^{12}$CO(1-0) and the $^{13}$CO(1-0) integrated intensities (dashed blue and solid red lines, respectively), superimposed on the $K_S$-band image. The OSO 20 m beam at 115~GHz is displayed in the left bottom corner. Labels are in units of K~km sec$^{-1}$. The white crosses indicate the positions observed.}
              \label{k12}%
%     \end{figure}         
    \end{figure*}
    
\subsubsection{CO and CS}

In the three maps of the CO transitions, all observed spectra clearly show line emission ($S/N > 5$).  In particular, the $^{12}$CO(1-0) spectra show a double-peak appearance, that can be easily interpreted as self-absorption after comparing with the peak positions of $^{13}$CO(1-0) and CS(2-1). In Figure~\ref{pos00}, the corresponding three spectra observed towards the (0,0) position (i.e. the IRAS source location) are overplotted, and it looks evident that the $^{13}$CO(1-0) and CS(2-1) peaks lie in the range where the $^{12}$CO(1-0) shows a dip between its two peaks. Therefore, as it is, this transition cannot be used to derive gas physical parameters 
but, on the other hand, indicates
a high column density cloud.

We first used the CS(2-1) line peak to derive the $V_\mathrm{LSR}$ of the cloud: the center of the Gaussian fit, in the (0,0) position, corresponds to $V_\mathrm{LSR}=-38.2$~km~s$^{-1}$. Similarly, for the $^{13}$CO(1-0), we obtained $V_\mathrm{LSR}=-38.0$~km~s$^{-1}$. Finally, fitting a Gaussian line profile to the $^{12}$CO(1-0) line wings \citep[cfr.][see Figure~\ref{pos00}]{kra04}, yields $V_\mathrm{LSR}=-38.5$~km~s$^{-1}$. Given the good agreement of all these velocities, we adopted the value $V_\mathrm{LSR}=-38.2$~km~s$^{-1}$.

According to the circular rotation model by \citet{bra93}, this value of $V_\mathrm{LSR}$ at the Galactic coordinates of this source corresponds to a heliocentric distance $d_H=5.5$~kpc, and a Galactocentric distance $d_G=13.6$~kpc. 
This is in good agreement with the distance quoted by \citet{pitann11,birkmann07}.
Thus, the projected sizes of the regions mapped in CO (both isotopes) and CS turn out to be $\sim4$~pc and $\sim2.4$~pc, respectively.

The maps of the integrated intensity $I=\int T_\mathrm{MB} \, dv$ of $^{12}$CO(1-0) and $^{13}$CO(1-0), obtained in the ranges between $-47$ and $-27$ km~s$^{-1}$ and $-45$ and $-30$ km~s$^{-1}$, respectively, reflect the different role of these tracers and the presence of saturated (self-absorbed) lines in the first one (see Figure~\ref{k12}). In fact, the $^{12}$CO(1-0) intensity appears arranged in a single ``clump'' peaked on the (0,0) position, whereas the $^{13}$CO(1-0) shows a further increase towards the west side of the map, corresponding to 
the second ``sub-cluster'' (labelled ``West'') that can be noticed in the infrared images (Figures~\ref{JHK}, \ref{WISE}).
The ``NW'' sub-cluster lies outside the range of the CO maps.

\subsubsection{Ammonia}
\begin{figure}
   \centering
   \includegraphics[width=8cm, angle=0, scale=0.8]{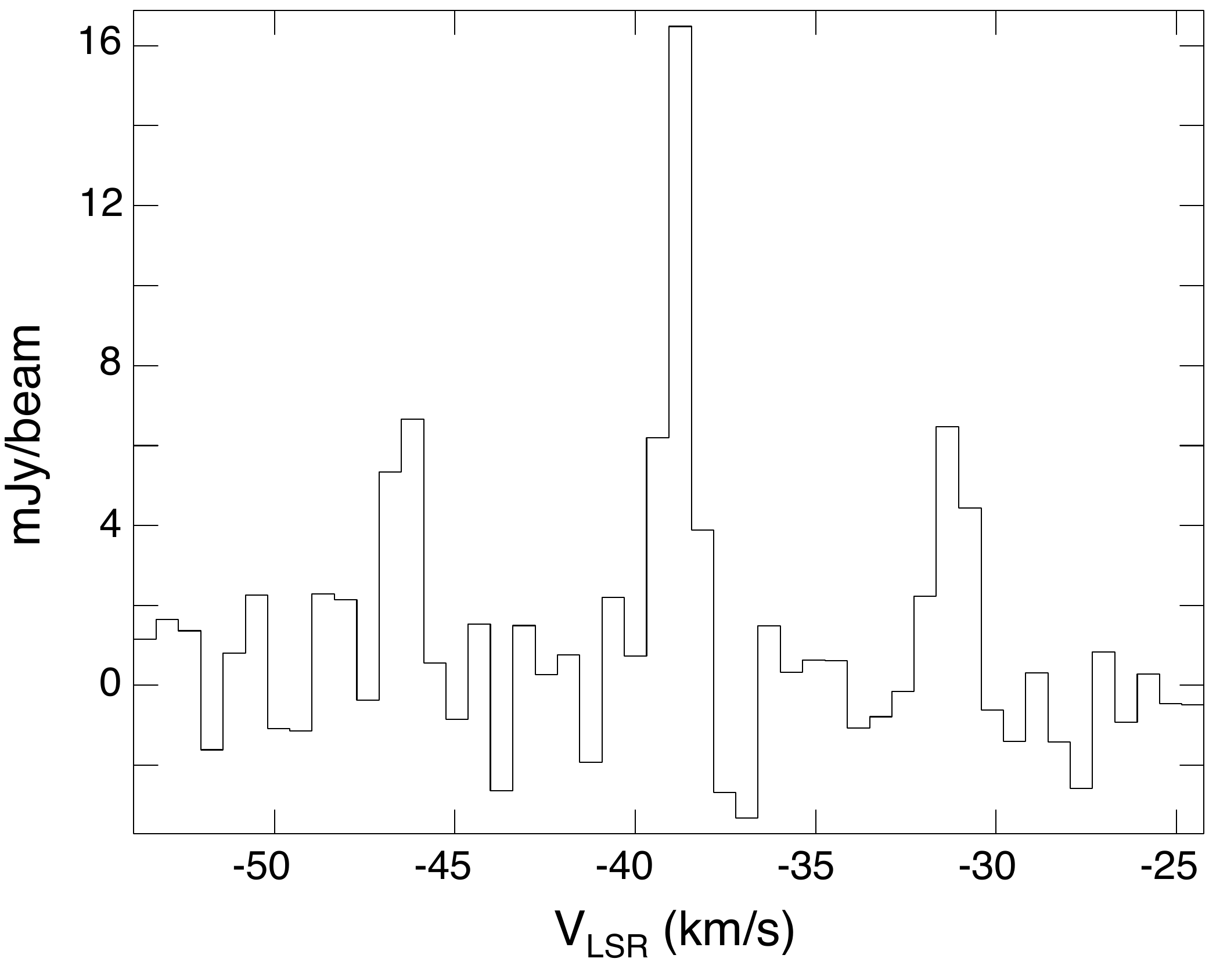}
   \caption{NH$_3$ (1,1) spectrum observed toward the peak position of the ammonia emission [$\alpha$(J2000) = 04$^h$22$^m$33.82$^s$,  $\delta$(J2000) = 51$^{\circ}$51$'$05.5$''$].}
              \label{ammonia1}%
     \end{figure}

We detected the main component and the inner satellite lines of the NH$_3$(1,1) transition in the velocity range $-$39.4~km~s$^{-1}$ $\lesssim$ V$_{LSR}$ $\lesssim$ $-$36.9~km~s$^{-1}$, with maximum intensity in the $-$38.8~km~s$^{-1}$ velocity channel, in very good agreement with the CO and CS data. NH$_3$(2,2) emission was not detected. In Figure~\ref{ammonia1} we show the observed NH$_3$(1,1) spectrum obtained toward the peak position of the emission.

\begin{figure}
   \centering
   \includegraphics[width=8cm, angle=0, scale=1.0]{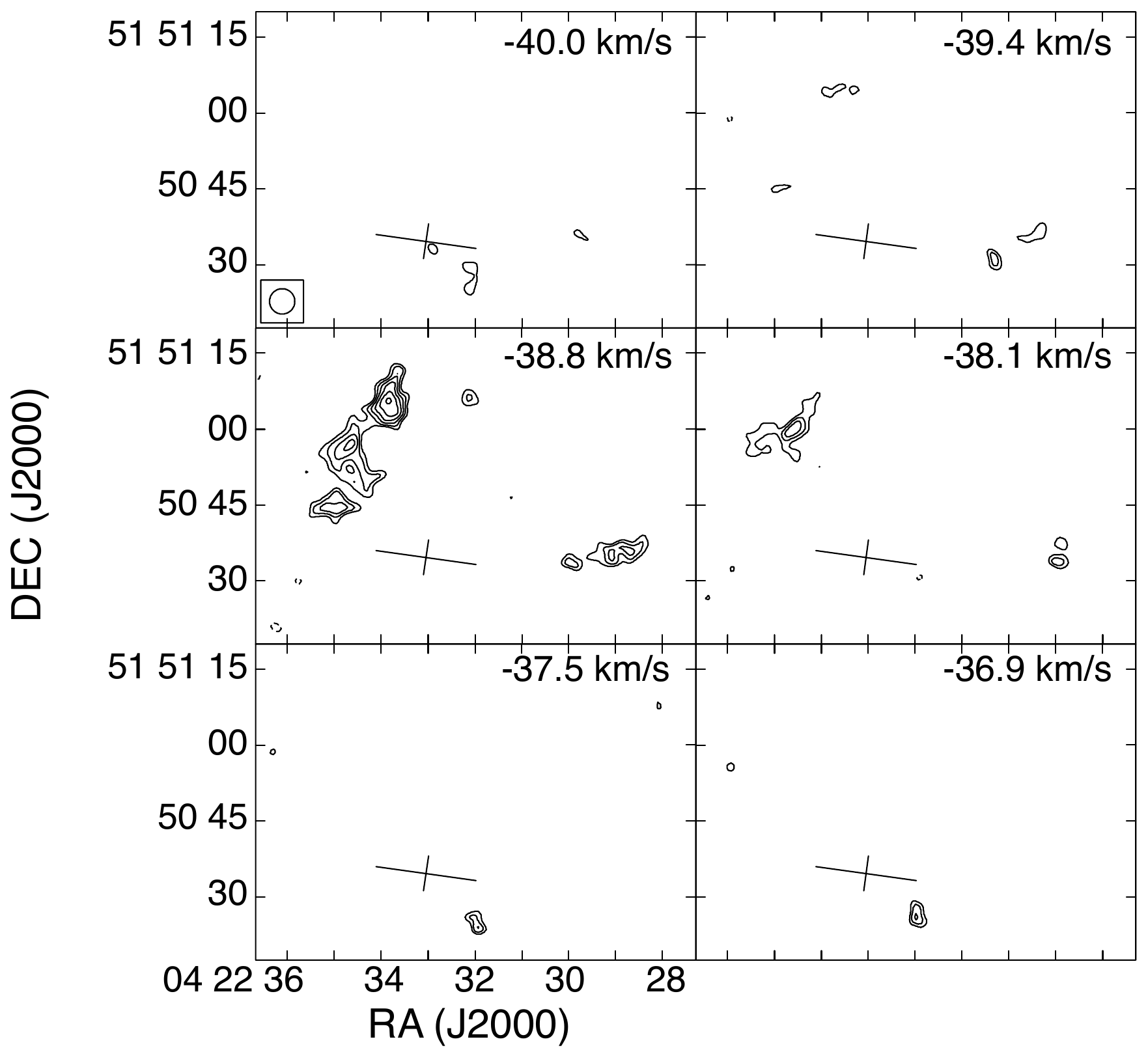}
   \caption{Velocity channel contour maps of the NH$_3$ (1,1) main line component towards the region of IRAS~04186+5143. The LSR velocity is indicated. Contours are -4, 4, 5, 6, 7, 8, 9 times 1.8~mJy~beam$^{-1}$, the rms noise of the maps. The synthesized beam is 5$''$. The nominal position of the IRAS source is shown by a cross (the size of the cross indicates its position error).}
              \label{ammonia2}%
     \end{figure}

In order to optimise the signal-to-noise ratio of the distribution of the NH$_3$(1,1) emission in the region, we have made images with natural weighting and a restoring beam of 5$''$. The corresponding contour maps of different velocity channels are shown in Figure~\ref{ammonia2}. These contour maps reveal that ammonia emission is not detected towards the IRAS source position where the main cluster is located. This is to be expected as objects embedded in ammonia cores are likely to be in earlier stages with little or no near/mid-infrared emission detectable. Instead,  high-density gas, commonly traced by ammonia, is seen here forming structures that could represent arcs around the cluster, possibly remaining gas from the original parent star-forming core.
In any case, the detection of ammonia at a V$_{LSR}$ coincident with that of CO corroborates the presence of Galactic dense molecular gas.

    \subsection{Young stars: the near-infrared view}
    
Table~\ref{JHKtable} (full table provided on-line only) gives the photometry of all NOTCam sources detected in the images. 

\begin{table}
\caption{NOT sources towards IRAS~04186+5143 (full table provided on-line)} 
\label{JHKtable}% title of Table
%\label{table:1}      % is used to refer this table in the text
\begin{tabular}{l c c c c c}
\hline\hline                 % inserts double horizontal lines
ID & R.A. & Dec & m${_J}$ & m${_H}$ & m$_{K\!s}$ \\% table heading
\# & $(^{\circ})$ & $(^{\circ})$ &  &  &  \\
\hline                        % inserts single horizontal line
\noalign{\vspace{2 pt}}
   1 &    65.58004 &    51.83786 &    17.45 &    16.12 &    15.30 \\
   2 &    65.58015 &    51.84943 &    19.41 &    18.47 &    18.27 \\
   3 &    65.58021 &    51.85655 &    18.66 &    17.27 &    16.61 \\
   4 &    65.58033 &    51.83777 &    17.68 &    16.46 &    15.52 \\
   5 &    65.58053 &    51.85194 &    17.62 &    16.98 &    16.93 \\
   6 &    65.58061 &    51.85741 &    19.58 &    18.67 &    18.51 \\
   7 &    65.58105 &    51.82590 &    20.26 &    19.11 &    18.23 \\
   8 &    65.58124 &    51.86619 &    19.79 &    18.79 &    18.58 \\
   9 &    65.58126 &    51.86396 &    $\ldots$ &    $\ldots$ &    19.10 \\
  10 &    65.58178 &    51.81469 &    $\ldots$ &    $\ldots$ &    19.36 \\
  11 &    65.58207 &    51.85682 &    15.21 &    14.82 &    14.62 \\
  12 &    65.58224 &    51.84251 &    20.52 &    18.62 &    17.48 \\
  13 &    65.58228 &    51.83663 &    $\ldots$ &    18.83 &    17.98 \\
  14 &    65.58233 &    51.87639 &    $\ldots$ &    $\ldots$ &    19.06 \\
  15 &    65.58234 &    51.87183 &    17.60 &    17.03 &    16.95 \\
  16 &    65.58256 &    51.85955 &    19.40 &    18.57 &    18.52 \\
   \hline\hline
%\noalign{\vspace{-10 pt}}
\end{tabular}
\end{table}

\begin{figure}
   \centering
   \includegraphics[width=8cm]{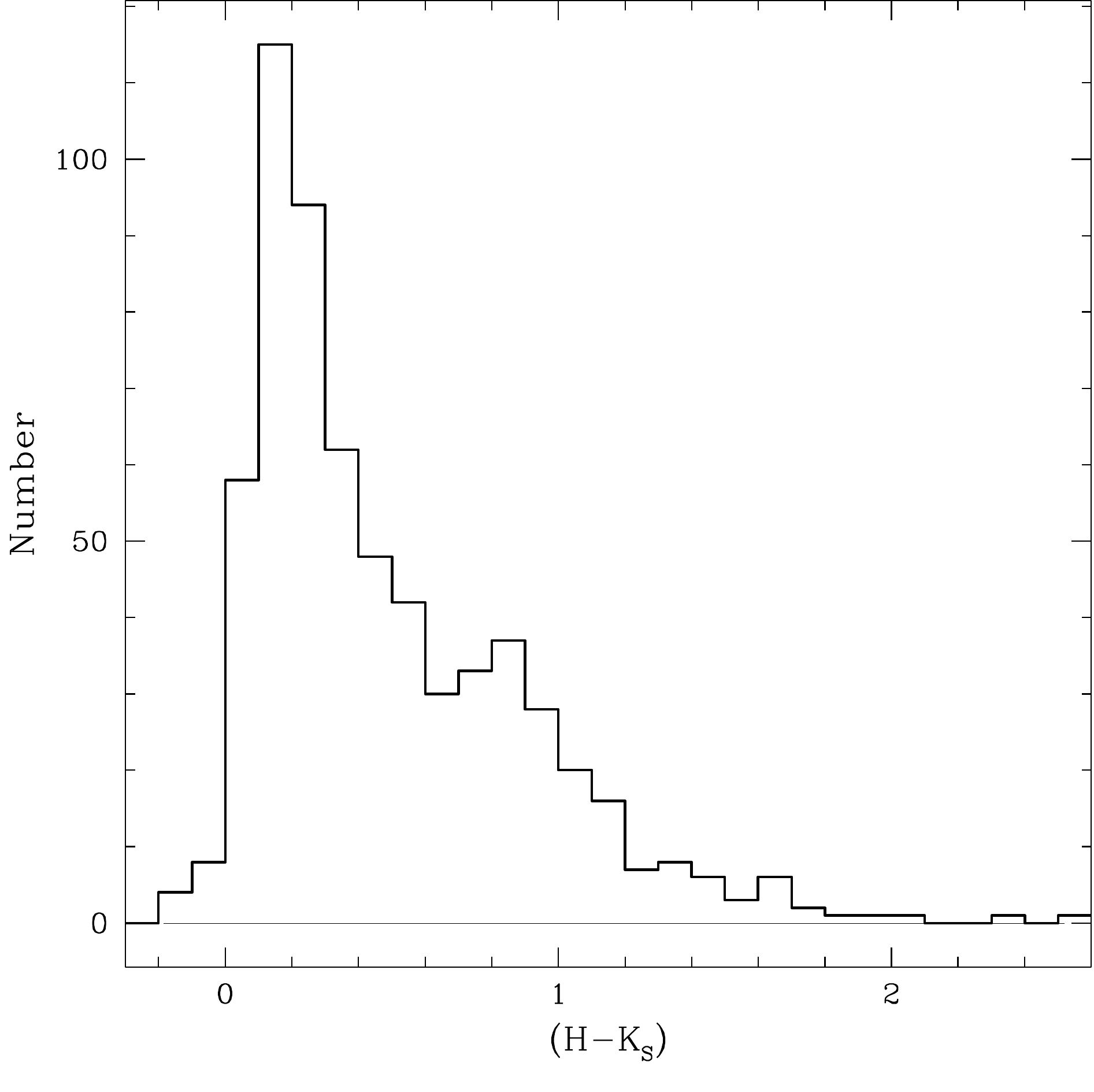}
   \caption{
Histogram of the observed $(H-K_S)$ colours. The well-defined peak is composed of foreground field sources (``blue'' sources). The red wing of the distribution is composed of sources spatially concentrated in the region where the molecular cloud is present (see Fig.~\ref{scatter}). 
	} 
	\label{histHK}%
    \end{figure}

Figure~\ref{histHK} shows the histogram for the observed $(H-K_S)$ colours of the sources detected in both the $H$ and the $K_S$-band images. 
The corresponding histogram of a normal star field (constituted by main-sequence stars and without the presence of embedded clusters) would be approximately a Gaussian, with the spread in $(H-K_S)$ values about the peak value stemming from the range of intrinsic colours of main-sequence stars and to low values of variable foreground extinction in the lines of sight of each source.
However, the histogram for the observed $(H-K_S)$ colours of the sources towards IRAS~04186+5143 shown in Fig.~\ref{histHK} clearly deviates from a Gaussian, exhibiting a red tail (with possibly a second peak) representing sources with large values of $(H-K_S)$. This observed excess near-IR emission must be due to the presence of embedded young stars.

\begin{figure}
   \centering
   \includegraphics[width=8cm]{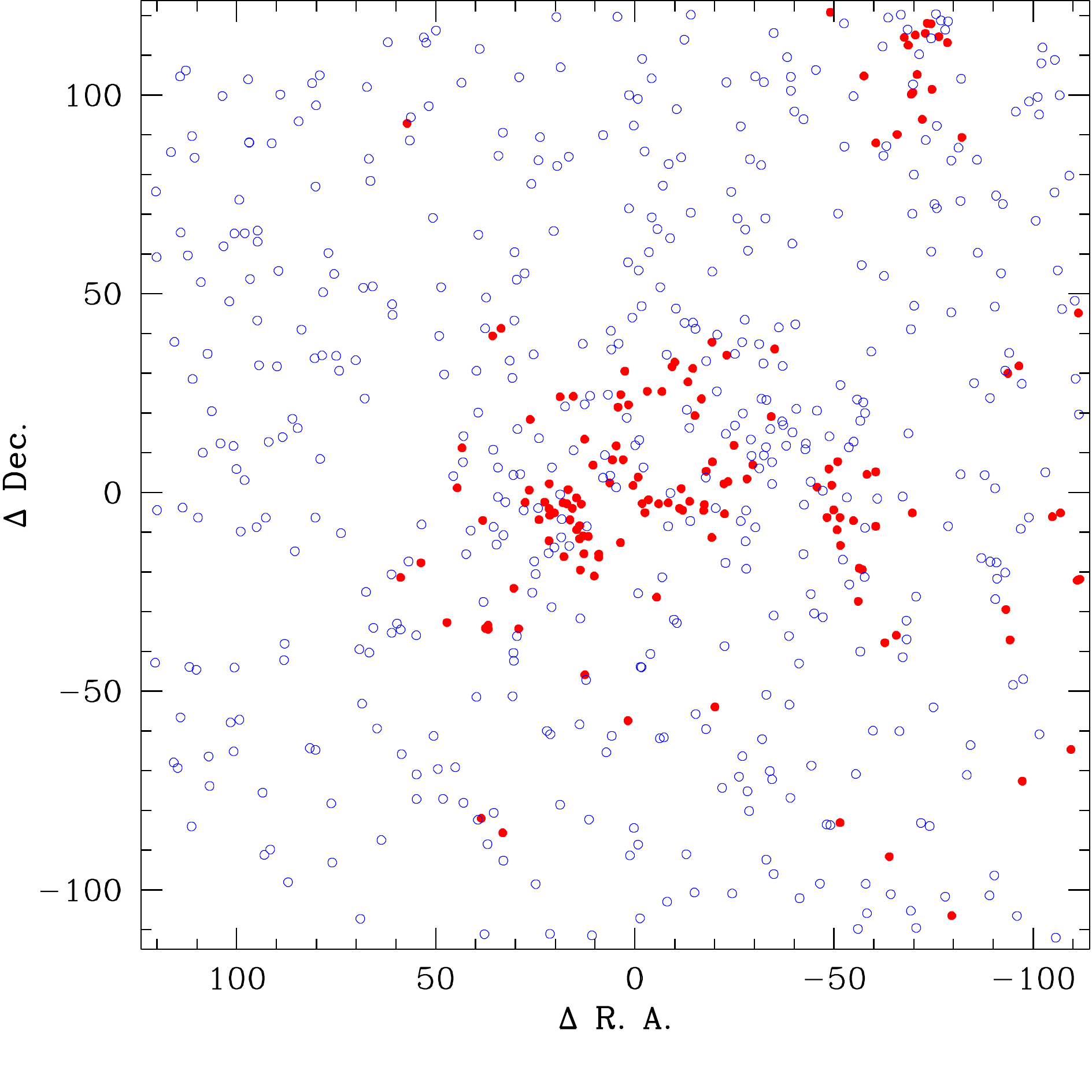}
   \caption{Spatial distribution of all the sources seen both in the $H$ and in the $K_S$ band images. Blue open circles represent sources with values of $(H-K) < 0.75$.  Red filled circles represent sources with values of $(H-K) \ge 0.75$.  This plot is centred on the $IRAS$ point source. }
   \label{scatter}
\end{figure}

It is instructive to check the spatial segregation of these sources by colour. 
A Gaussian fit around the peak, excluding the red tail sources, yields a mean of $(H-K_S)=0.20$, with a standard deviation of 0.22 for the ``blue'' peak main-sequence field stars.
The spatial location of these peak ``blue'' sources is seen in Fig.~\ref{scatter} where they are represented by blue open circles. Red filled circles, on the other hand, represent sources in the red tail of the $(H-K_S)$ histogram, the ``red'' sources with $(H-K_S) > 0.75$, that is $2.5\sigma$ away from the blue peak. The blue sources are scattered randomly and uniformly across the image, whereas the red sources are concentrated in the region of the molecular cloud. Taken together, these results strongly indicate that most red sources are objects associated with the cloud, either located behind the cloud (very few sources given the location of the cloud in the far outer Galaxy), or being embedded in the cloud and possibly containing near-infrared excess emission from circumstellar material. 

%______________________________________________ 
%
%______________________________________________________________

Blue sources, on the other hand, may be composed of a mix of foreground field sources and YSOs in a more evolved evolutionary stage. These more evolved young stars, if present, could be pre-main-sequence objects or even intermediate-mass or massive main-sequence stars formed in this cloud, which evolve much faster than their lower-mass siblings formed at the same time. Their higher masses would also contribute to their being bluer and thus not being told apart by red colours.

\bigskip

%______________________________________________ 
   \begin{figure}
   \centering
   \includegraphics[width=8cm]{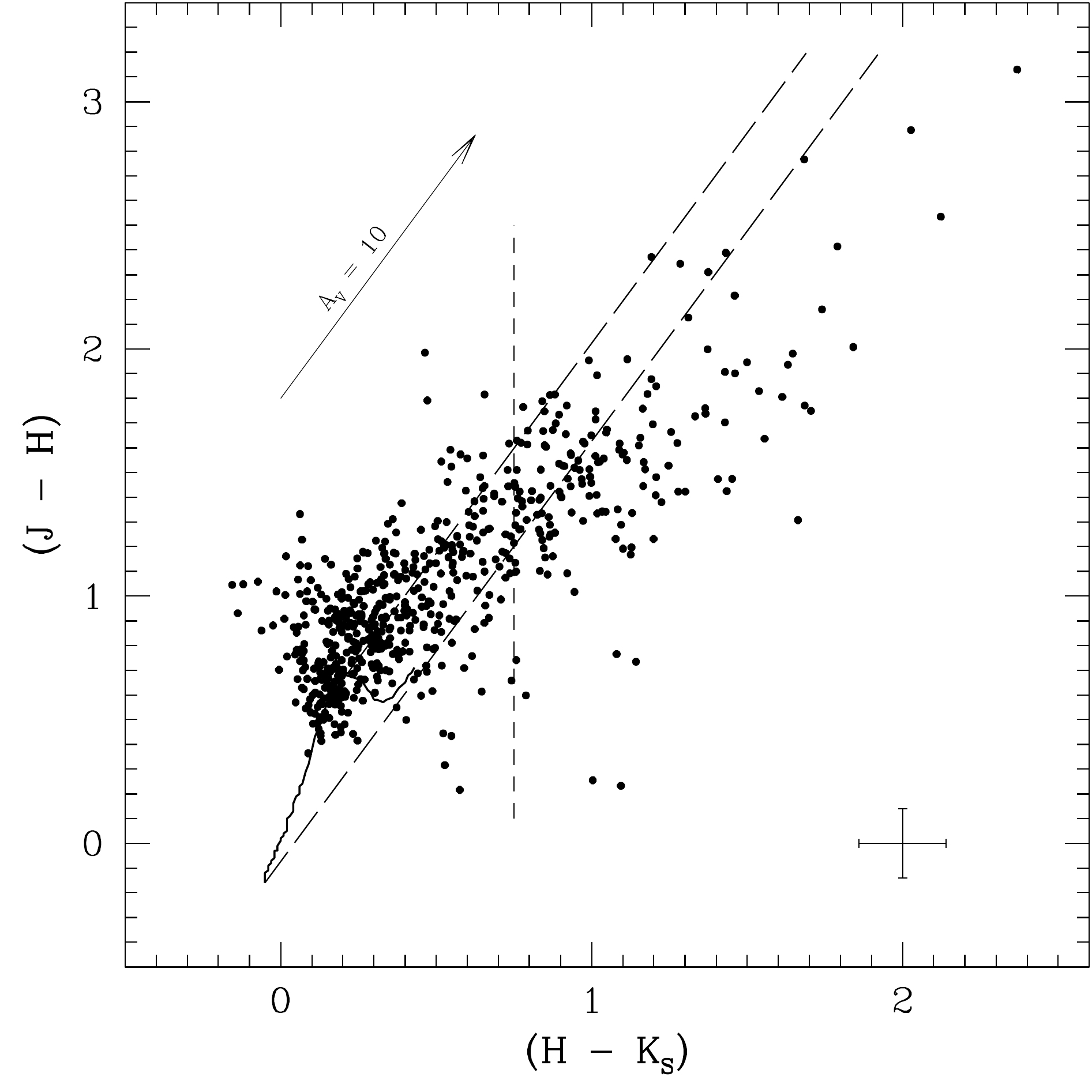}
   \caption{
Near-infrared colour-colour diagram of the region towards IRAS~04186+5143. 
The solid line represents the loci of unreddened main-sequence stars \citep{bessell88}, while long-dashed lines indicate the reddening band. The reddening vector indicates the direction of the shift produced by extinction by dust with standard properties.
The location of the vertical dashed line, derived from Fig.\ref{histHK}, is at $(H-K_S) = 0.75$. The cross in the lower right corner represents a typical error bar.
	}
	\label{cc}%
    \end{figure}
%
%______________________________________________________________

Using the point sources detected in all three $J$, $H$, and $K_S$-bands, 
we plotted the near-infrared colour-colour diagram, $(J-H)$ versus $(H-K_S)$, shown in Fig.\ref{cc}.
Most stars are located within the reddening band where stars appear if they are main-sequence stars reddened according to the interstellar extinction law \citep{rieke85}, which defines the reddening vector (traced here for $A_V=10$).
Pre-main-sequence YSOs, or massive main-sequence stars recently formed in this region, which have had time to clear the inner regions of their circumstellar discs, lie in this region as well. Giant stars appear slightly above this band. On the other hand, the location of stars to the right of the reddening band cannot be the result of interstellar reddening alone. They require the effect of emission by hot dust
 such as that in thick circumstellar discs or envelope molecular cloud cores. Thus, they are likely to be embedded young star objects with infrared excess emission from circumstellar material \citep{adams87}. 

For the sources that lie inside the reddening band, the highest value of $(H-K_S)$ is about 1.4. Using the mean value of $(H-K_S) \sim 0.3$ for field stars (according to Fig.~\ref{histHK}), we obtain a colour excess $E(H-K_S) = 1.1$ due to intra-cloud extinction. This value corresponds to a maximum visual extinction produced by the cloud core, through lines-of-sight where stars can be detected, of $A_V \sim 17$ \citep{rieke85}.

The location of the vertical dashed line, derived from Fig.~\ref{histHK}, is at $(H-K_S) = 0.75$. The two groups of sources, blue sources with $(H-K_S) < 0.75$ and red sources with $(H-K_S) \ge 0.75$, are very differently distributed on the colour-colour diagram.  A large fraction of the red sources are located outside and to the right of the reddening band, whereas the blue sources mostly occupy the inside of the reddening band. Thus, most red sources are likely to be YSOs. Given their spatial concentration (Fig.~\ref{scatter}), these red sources with $(H-K_S) \ge 0.75$ together with a fraction of the blue sources in this region, seem to represent a small young embedded stellar cluster of about 100 young stars forming in the molecular cloud. 
The actual number of stars in this young cluster is likely to be larger for at least three reasons. Firstly, we chose a conservative value of $(H-K_S) = 0.75$ ($2.5\sigma$ above the mean value of the $(H-K_S)$ of field main-sequence stars). Secondly, there are some young stars in a more advanced stage of the star formation process, already free of circumstellar material and exhibiting blue colours, thus not pinpointed by our colour selection criterion. Thirdly, we have considered only stars detected in all three $JHK_S$ bands.

 We can make a rough estimate of the total mass present in the stellar content of this cluster. A first estimate results from assuming 1 $M_\odot$ stars, yielding about 100 $M_\odot$. Not much different values are obtained, {\it e.g.} adopting a Salpeter Initial Mass Function (IMF) and a reasonable range of masses: the result is a total stellar mass of about 140 $M_\odot$.

We derive an upper limit for the masses of the YSOs present in this cluster in the following mode. The luminosity from the cluster region is dominated by the mid and far-infrared flux as measured by IRAS. We estimate this $L_{\rm FIR}$ to be about $4.6\times 10^3 L_{\odot}$. Assuming 
that all this luminosity is produced by a single star, this would set an upper limit of about 9--10  M$_{\odot}$ for any massive star present in this cluster. We conclude that the young stellar population present in this region is composed of low and intermediate-mass stars.

\subsection{The {\it Spitzer} view}

\begin{table*} 
\caption{Photometric data of Spitzer sources towards IRAS~04186+5143 having NOTCam counterparts (full table provided on-line) }  % title of Table
\label{spitzer-notcam}
\begin{tabular}{l c c c c c c c c c c c c }
%\label{table:1}      % is used to refer this table in the text
\hline\hline                 % inserts double horizontal lines
& &  &  & & &  \multicolumn{4}{c}{Flux} &  &  & \\
\cline{7-10}
ID & R.A.  & Dec & m${_J}$ & m${_H}$ & m$_{K\!s}$ & $[3.6]$  & $[4.5]$  & $[5.8]$  & $[8.0]$  & $[3.6]-[4.5]$ & $[5.8]-[8.0]$ & YSO  \\% table heading
\# & $(^{\circ})$ & $(^{\circ})$ & &  &  & (Jy) & (Jy) & (Jy) & (Jy) & & & class \\
\hline                        % inserts single horizontal line
\noalign{\vspace{2 pt}}
%   1 &   65.58008 &    51.83772 &    17.45 &    16.12 &    15.30 &  $\ldots$ &  8.216E-04 &  1.205E-03 &  1.258E-03 &    $\ldots$ &     0.68 & $\ldots$ \\
   2 &   65.58012 &    51.85640 &    18.66 &    17.27 &    16.61 &  8.861E-05 &  6.169E-05 &  $\ldots$ &  $\ldots$ &     0.09 &    $\ldots$ & $\ldots$ \\
   3 &   65.58038 &    51.83772 &    17.45 &    16.12 &    15.30 &  8.344E-04 &  1.130E-03 &  1.205E-03 &  1.258E-03 &     0.81 &     0.68 & II    \\
   4 &   65.58049 &    51.85184 &    17.62 &    16.98 &    16.93 &  4.274E-05 &  2.621E-05 &  $\ldots$ &  $\ldots$ &    -0.05 &    $\ldots$ & $\ldots$ \\
   5 &   65.58195 &    51.85678 &    15.21 &    14.82 &    14.62 &  3.358E-04 &  1.990E-04 &  1.330E-04 &  4.850E-05 &    -0.08 &    -0.46 & $\ldots$ \\
   6 &   65.58226 &    51.87177 &    17.60 &    17.03 &    16.95 &  4.411E-05 &  2.499E-05 &  $\ldots$ &  $\ldots$ &    -0.13 &    $\ldots$ & $\ldots$ \\
   7 &   65.58245 &    51.84241 &    20.52 &    18.62 &    17.48 &  1.971E-04 &  2.828E-04 &  3.685E-04 &  4.912E-04 &     0.88 &     0.95 & I     \\
   8 &   65.58312 &    51.81478 &    $\ldots$ &    18.87 &    18.13 &  3.614E-05 &  2.881E-05 &  $\ldots$ &  $\ldots$ &     0.24 &    $\ldots$ & $\ldots$ \\
   9 &   65.58325 &    51.81277 &    17.63 &    16.83 &    16.58 &  6.500E-05 &  4.055E-05 &  $\ldots$ &  $\ldots$ &    -0.03 &    $\ldots$ & $\ldots$ \\
  10 &   65.58411 &    51.84328 &    $\ldots$ &    $\ldots$ &    17.79 &  8.834E-04 &  1.419E-03 &  2.952E-03 &  1.952E-03 &     1.00 &     0.19 & I     \\
  11 &   65.58423 &    51.84535 &    16.92 &    16.13 &    15.90 &  1.154E-04 &  7.445E-05 &  $\ldots$ &  $\ldots$ &     0.01 &    $\ldots$ & $\ldots$ \\
  12 &   65.58426 &    51.87494 &    18.98 &    17.84 &    17.30 &  8.294E-05 &  8.093E-05 &  7.059E-05 &  6.755E-05 &     0.46 &     0.59 & II    \\
  13 &   65.58432 &    51.87392 &    17.74 &    16.63 &    16.35 &  1.045E-04 &  8.625E-05 &  7.429E-05 &  7.952E-05 &     0.28 &     0.71 & II    \\
  14 &   65.58480 &    51.87159 &    17.35 &    16.54 &    16.36 &  1.871E-04 &  $\ldots$ &  $\ldots$ &  $\ldots$ &    $\ldots$ &    $\ldots$ & $\ldots$ \\
\hline\hline
%\noalign{\vspace{-10 pt}}
\end{tabular}
\end{table*}
%\end{landscape}

\begin{figure}
   \centering
      \includegraphics[width=8cm,angle=0]{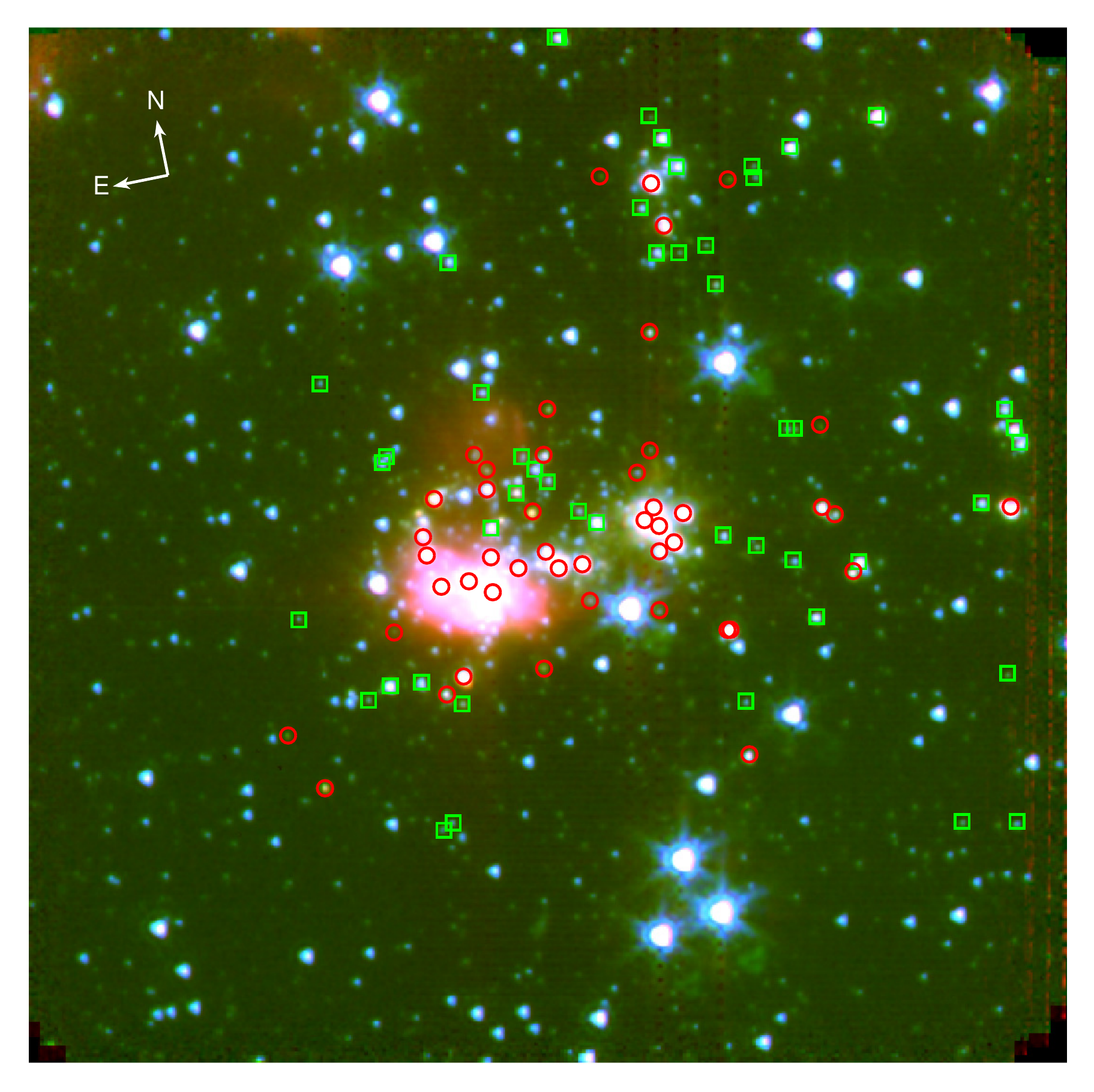}
   \caption{
{\it Spitzer} 3.6 (blue), 4.5 (green), and 5.8 (red) $\mu$m colour composite
image towards IRAS 04186+5143 covering about 5.7' $\times$ 5.7' with
Class~I sources (red circles) and Class~II sources (green squares) overlaid.
	} 
	\label{spitzer}%
    \end{figure}

Figure~\ref{spitzer} presents the {\it Spitzer} bands 1-2-3 colour composite image obtained towards IRAS~04186+5143. As expected, a clear concentration of stars is seen close to the centre of the image. These mid-infrared counterparts of the $JHK_S$ sources appear quite ``red'' and support the idea that we are dealing with a young stellar cluster located far in the outer Galaxy. 

%______________________________________________ 
%
%_____________________________________________________________

\begin{figure}
   \centering
      \includegraphics[width=6.5cm,angle=270]{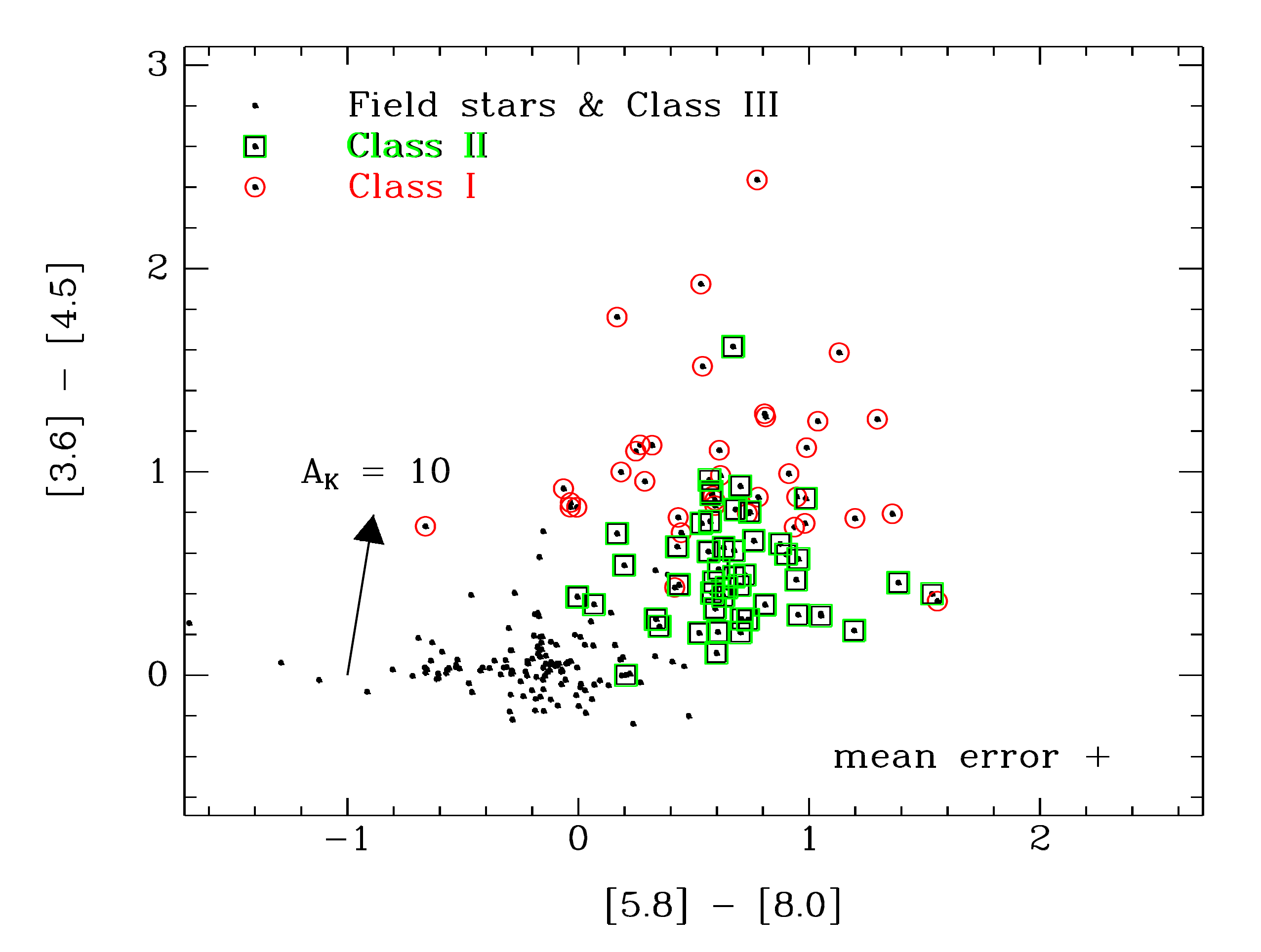}

   \caption{
{\it Spitzer} colour-colour diagram of the region towards IRAS~04186+5143. 
	} 
	\label{colcol}%
    \end{figure}

In Figure~\ref{colcol}, we present the {\it Spitzer} four-band colour-colour diagram of this region.
Following YSO classification criteria, e.g. \citet{allen04, gutermuth08, gutermuth09}, 
we used different coloured-symbols to indicate different types of sources: class~I (red circles), class~II (green squares), and class~III and field stars (black dots). 
Our sample is restricted to those sources with photometric errors less than 0.2 mag
in all four IRAC bands. We first select as Class~I sources those that either have
$([4.5]-[5.8] > 1)$ or $([4.5]-[5.8] > 0.7$ and $[3.6]-[4.5] > 0.7)$, referring here to
colour indices in magnitudes. From the remaining sources in the sample, the Class~II
sources are those which fulfill the three requirements:
$([4.5]-[8.0] > 0.5)$ and $([3.6]-[5.8] > 0.35)$ and $([3.6]-[5.8] \le (3.5\times (([4.5]-[8.0])-0.5)+0.5)$.
This is very close to \citet{gutermuth08, gutermuth09} except that we have ignored the
potential extragalactic contaminants which should be of marginal importance in the
small area explored here.
We find 37 Class I and 48 Class II objects in the sample of 215 sources with photometry in all four IRAC bands. 
The large fraction of Class I sources is a clear sign of an active and very young
region. A typical cluster core has a Class II / Class I ratio of 3.7 according to the
large survey by \citet{gutermuth09}. 
The presence of a fair number of Class~I sources that are usually associated with jets or outflows may explain the detection by \citet{pitann11} of diffuse H$_2$, [Si~II], [Fe~II], and [Ne~II] {\it Spitzer} spectral lines that may indicate the presence of shock-excited gas.

%______________________________________________ 
%   \begin{figure*}
%
%_____________________________________________________________

\begin{figure}
   \centering
      \includegraphics[width=6.5cm,angle=-90]{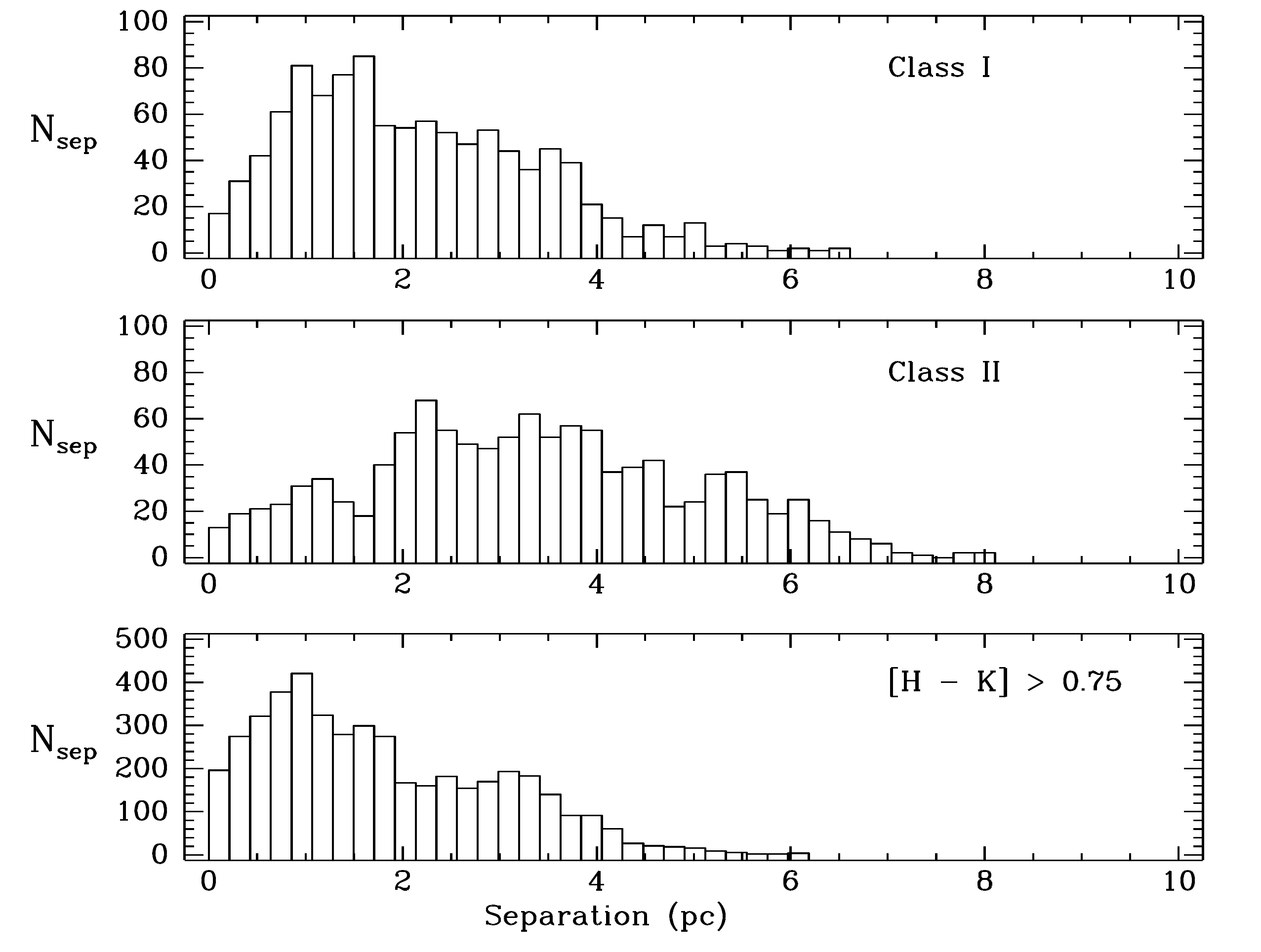}
   \caption{
The distribution of projected separations between sources in the different groups:
Class I sources (top), Class II sources (middle), NIR sources with $[H-K] > 0.75$
(bottom). The number of separations per bin is shown versus projected separation
(in pc) assuming a distance of 5.5 kpc, using a bin size of $8"$ ($\sim$~0.2 pc).
	} 
	\label{histos}%
    \end{figure}

An inspection of the spatial distribution of the Class~I and Class~II sources (see Fig.~\ref{spitzer}) reveals that they are all
predominantly found near the dense clumps detected by {\it Herschel} (see below), with a clear sub-clustering of at least the Class I sources. 
In Figure~\ref{histos} we show a simple representation of their spatial distribution through
the histogram of all projected separations between sources within the same group \citep{kaas04}. A homogeneous distribution would
give a broad gaussian, while clustering shows up as structure, where peaks relate to the
clustering scale. A binsize of $8"$ (0.2 pc) was used to ensure a sufficient population
within the smallest bin. 
In order to test the stability and the statistical significance of the peaks in the
distribution, we have varied the binsize in steps from about half to about 1.5 times
this value.
The strongest peak gives the approximate diameter of the most populous group. 
For the Class~I sample there is an indication of two separate peaks at small source
separations. The statistical significance is only about 1~sigma in this histogram,
however, and with larger binsizes the two peaks merge to a broad maxima across the
range 0.8 -- 1.7~pc, which is significant to $>$ 4~sigma.
The Class~II population has a relatively minor peak at 1.2~pc while it has its main
peak at 2.2~pc. This clearly shows a stronger clustering for Class~Is than for Class~IIs.
The sample of near-IR sources with $[H-K] > 0.75$ magnitudes has two peaks at the same
location as the Class~I sources. Lowering the binsize to $4''$ for this more numerous
sample, we fine-tune the locations of these peaks to 0.9~pc and 1.7~pc with a high
statistical significance. In addition, this sample has a broader feature at 3 pc
reflecting the NW sub-cluster distance to the main clusters. 
This latter population is expected to include both the Class~I and Class~IIs and many more sources not resolved/detected with IRAC. The fact that it follows so well the small scale structure of the Class~Is, suggests that most of these sources are likely very young cluster members.

%______________________________________________ 
%   \begin{figure*}
%
%_____________________________________________________________

Table~\ref{spitzer-notcam} (full table provided on-line only) contains the photometry of all Spitzer sources having $J$, $H$, or $K_S$ counterparts detected in our NOTCam near-IR images, with indication of YSO classification when available.

\subsection{The Herschel view}

\begin{figure*}
   \centering
   \includegraphics[width=16cm]{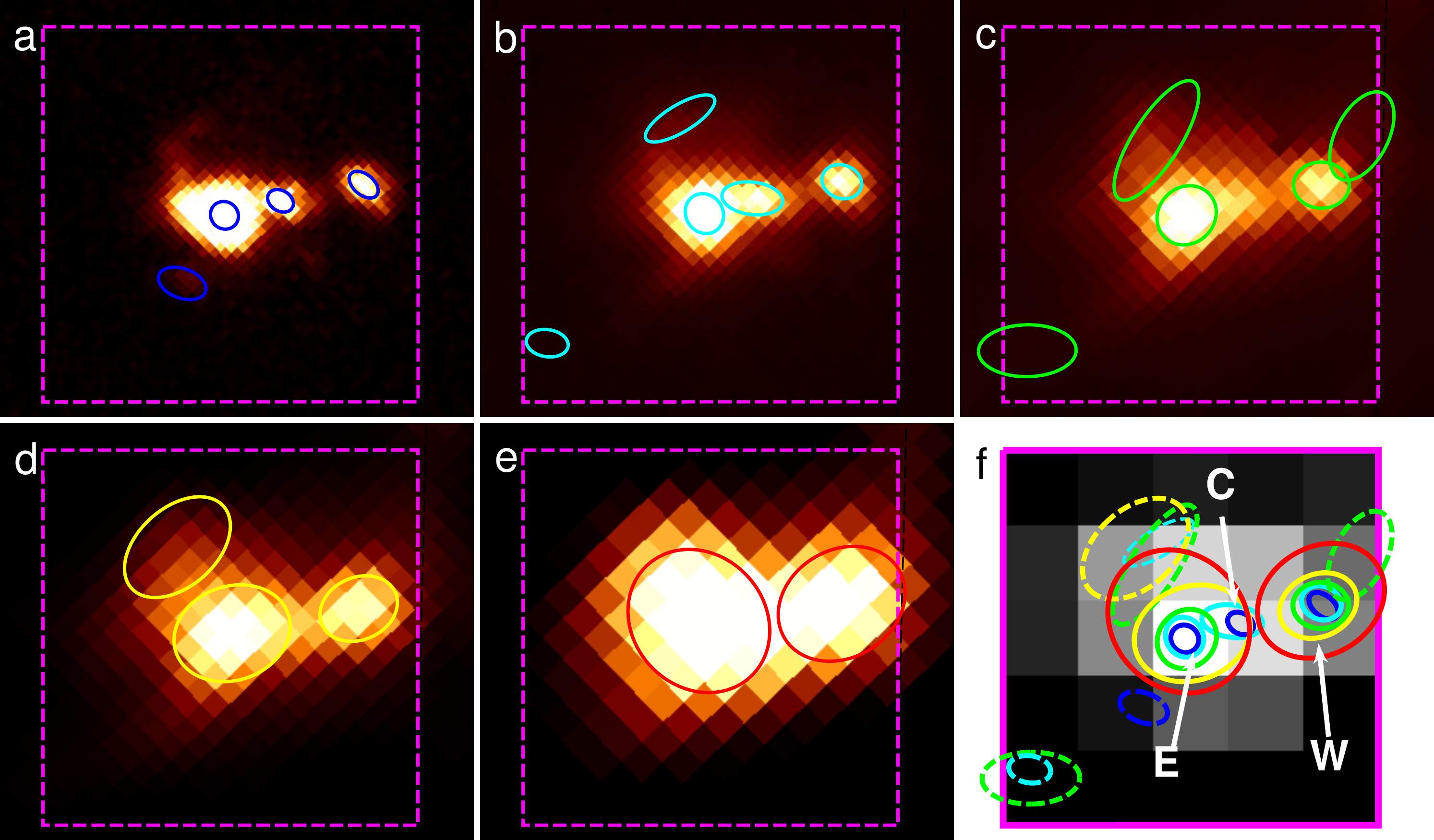}
   \caption{\textit{a}, \textit{b}, \textit{c}, \textit{d}, \textit{e}: Hi-GAL maps of the region around
   IRAS~04186+5143 at 70, 160, 250, 350, and 500 $\mu$m, respectively, covering the area mapped in CO (magenta dashed box). The ellipses denote the FWHM of the sources extracted by CuTEx at the
   various wavelengths. \textit{f}: CO(1-0) integrated intensity in grey scale, with the sources
   identified in panels \textit{a}-\textit{b} superposed (using the same color-wavelength convention of previous panels). Solid ellipses
   identify the counterparts of the three sources having a reliable SED (see text), indicated with ``E'', ``C'', and ``W'', respectively, whereas dashed ellipses identify sources detected in only one or two bands, and discarded for further analysis.}
   \label{6frames}
\end{figure*}

In this section we present and discuss the photometric data of 
Hi-GAL sources detected as described in Sect. 2.4.
Figure~\ref{6frames} presents the Hi-GAL maps of the region surveyed in the CO transitions, at 70, 160, 250, 350, and 500 $\mu$m, respectively.

\begin{figure*}
   \centering
   \includegraphics[width=13cm]{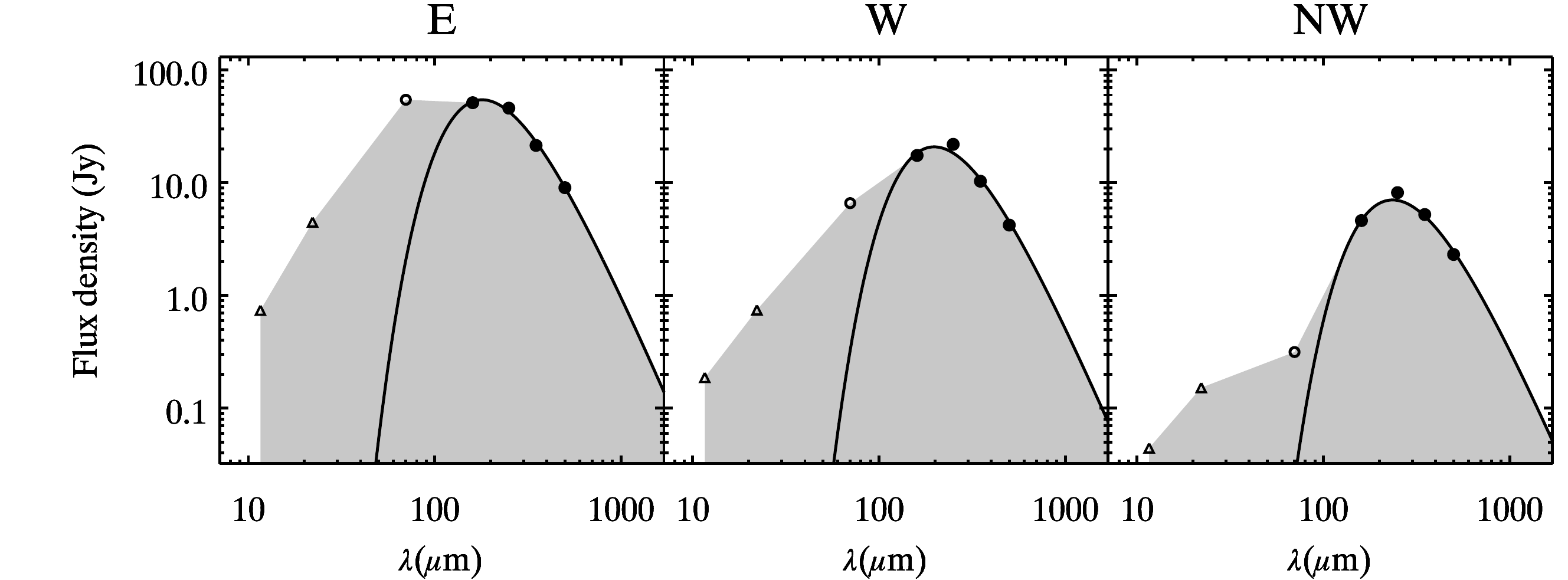}
   \caption{Mid and far-infrared spectral energy distributions of the three main Hi-GAL sources found in the investigated region, the source ``E'' being coincident with the location of IRAS 04186+5143. WISE fluxes are represented with triangles, while Herschel ones are marked with circles. Fluxes taken into account for the modified grey body fit (shown with a solid curve) are plotted with filled symbols. The grey-shaded area represents the integral computed to determine the bolometric luminosity $L_\mathrm{bol}$.
   }
   \label{seds}
\end{figure*}

Two sources (clumps) were clearly detected at all five bands (see panel \textit{f}) : the eastern one (E) clearly corresponds to the CO peak and to the main cluster location, while the western one (W) is associated with the second cluster.
A few further and fainter detections, at only one or two bands, represent less reliable sources, with the exception of the one located midway between E and W, which we designate by source C, and which coincides with a further overdensity of sources in the $JHK_S$ images. In this case, indeed, the source C is clearly visible in the 70~$\mu$m and 160~$\mu$m maps, getting confused with E longward of 160~$\mu$m. Given the limited spectral coverage for C, it is not possible to estimate the physical conditions of its envelope through a best-fit procedure based on a modified blackbody model, whereas thanks to the availability of SPIRE fluxes this procedure is feasible for the E and W clumps (Figure~\ref{seds}).

Before performing this fit, we note that the beam-deconvolved observed sizes $\theta_{\lambda}$
noticeably increase at the Herschel 350 and 500~$\mu$m bands, so that the fluxes measured at these
wavelengths come from larger volumes of dust \citep{mot10,gia12}. Thus, we adopted a flux scaling strategy, following that of \citet{eli10}, where we impose 
$\bar{F_{\lambda}}=F_{\lambda}\times \left(\theta_{250}/\theta_{\lambda} \right)$, for $\lambda=350,500~\mu$m.
This is based on the assumptions that (i) the source is optically thin at $\lambda \geq 250~\mu$m
(which is used as reference wavelength), (ii) the temperature gradient is weak \citep{mot01}, and (iii) the radial
density profile is $\rho(r) \propto r^{-2}$ (i.e. $M(r) \propto r$),

We then fitted a modified black body to the four fluxes from 160 to 500~$\mu$m. Since the 70~$\mu$m flux generally
shows an excess due to the proto-stellar content of the clump \citep{sch12}, we keep it as an upper limit to
further constrain the fit. The modified black body expression is:
\begin{equation}\label{gb}
F_{\nu}=(1-e^{-\tau_{\nu}})B_{\nu}(T_d)\Omega~,
\end{equation}
where $F_{\nu}$ is the observed flux density at the frequency
$\nu$, $B_{\nu}(T_d)$ is the Planck function at the dust temperature $T_d$, and $\Omega$ is the source
solid angle in the sky. The optical depth is given by
\begin{equation}\label{tau}
\tau_{\nu}=(\nu/\nu_0)^\beta~,
\end{equation}
where $\nu_0=c/\lambda_0$ is the frequency at which $\tau=1$, and $\beta$ is the exponent of the
power-law dust emissivity at large wavelengths. 
Four free parameters are present in the previous equations. In order to reduce their number, we imposed $\beta = 2$, as in \citet{eli13} \citep[see][for a detailed justification of this choice]{sad13}, and $\Omega$ to be equal to the source area observed at 250~$\mu$m. Applying this procedure, only $T$ and $\lambda_0$ are left free to change\footnote{Choosing lower values for $\beta$, e.g. $\beta = 1.5$, can lead to different values of clump masses, as in the case of clumps E and W whose masses are found to be smaller by a factor $\sim 2$ compared to those below, but with a much worse $\chi^2$.}.

The clump mass is subsequently derived from
\begin{equation}\label{mthick}
M=(d^2\Omega/k_\mathrm{ref})\tau_\mathrm{ref}~,
\end{equation}
\citep[cf.][]{pez12}, where $k_\mathrm{ref}$ and $\tau_\mathrm{ref}$ are the opacity and the optical depth, respectively,
estimated at a given reference wavelength $\lambda_\mathrm{ref}$. Here we chose $k_\mathrm{ref}=0.1$~cm$^2$~g$^{-1}$ at
$\lambda_\mathrm{ref}=250~\mu$m \citep[][which already accounts for a gas-to-dust ratio of 100]{hil83},
while $\tau_\mathrm{ref}$ is obtained from Equation~\ref{tau}.

The physical properties of the clumps E and W are reported in Table~\ref{hitab}. We remark that these refer
to the volume enclosed within the source size observed at 250~$\mu$m, which is $15.5\arcsec$ and $9.7\arcsec$
for the clumps E and W, respectively, well below the CO map grid step. The coordinates given are those of the counterparts found at 70 $\mu$m (namely the shortest wavelength available).

\begin{table*}
\caption{Photometric and physical properties of the main Hi-GAL sources}
\label{hitab}
\renewcommand{\footnoterule}{}  % to avoid a line before footnotes
\begin{minipage}{16cm}
\begin{tabular}{lcccccccccccc}
\hline\hline
Designation & $\alpha$ & $\delta$ & $F_{70}$ & $F_{160}$ & $F_{250}$ & $F_{350}$ & $F_{500}$ & $\theta_{250}$ &  $M$ & $T$ & $\lambda_0$ & $L_\mathrm{bol}$ \\ 
 &      deg & deg      &   Jy     & Jy        & Jy        & Jy        & Jy        & $\arcsec$      &  $M_{\odot}$ & K & $\mu$m & $L_{\odot}$ \\ 
\hline
E & 65.6354 & 51.8417 & 54.4 & 51.2 & 45.9 & 21.4 & 9.0 & 15.5 & 719 & 17.1 & 83.9 & 3150 \\
C & 65.6252 & 51.8435 & 10.0 & 18.1 & $\ldots$  & $\ldots$  & $\ldots$ & $\ldots$  & $\ldots$  & $\ldots$  & $\ldots$  &   $>510$ \\
W & 65.6105 & 51.8454 & 6.6 & 17.4 & 21.9 & 10.3 & 4.2 & 9.7 & 435 & 15.7 & 104.0 & 640 \\
NW & 65.6000 & 51.8758 & 0.3 & 4.6 & 8.2 & 5.2 & 2.3 & 25.8 & 416 & 12.4 & 38 & 125 \\
\hline
\end{tabular}
\end{minipage}
\end{table*}

Since the clump E is well-contained in our emission line maps, a comparison with CO-derived masses is possible provided we consider only the central pointing of the
CO maps. Two methods can be exploited to compute mass estimates from our CO observations, as done, for example,
in \citet{yun09}. Here, however, we want to consider the Gaussian profiles fitted to the wings of the self-absorbed
observed lines as genuine recovered line profiles.

The first method is based on the empirical linear relation between column density and the $^{12}$CO(1$-$0)
integrated intensity, $N(H_2)=X_{CO} I_{CO}$, with $X_{CO}$ larger in the outer Galaxy than in the nearby star 
formation regions ($X_{CO} \sim 2 \times 10^{20}$~cm$^{-2}$~K$^{-1}$~km$^{-1}$~s), given by \citet{nak06}:
$X_{CO}[$cm$^{-2}$~K$^{−1}$~km$^{-1}$~s$]=1.4\times 10^{20} \exp(R/ 11$~kpc), so that in our case
$X_{CO}=4.8\times 10^{20}$~cm$^{-2}$~K$^{-1}$~km$^{-1}$~s. The mass at the central pixel obtained using this method is $M_X=716~M_{\odot}$.

The second method assumes local thermal equilibrium (LTE) conditions, using $^{12}$CO(1$-$0) as an optically
thick line and $^{13}$CO(1$-$0) as an optically thin line \citep[see, e.g.][]{pin08}. The excitation temperature
is extracted from the peak main beam temperature of the CO(1$-$0) ($T_\mathrm{ex}= 31$~K). Assuming that excitation
temperatures are the same for both lines, the column density of $^{13}$CO is calculated through LTE relations,
(e.g. equations 6 and 4 of \citet{bra95}, respectively). In order to obtain H$_2$ column densities, a
$\left[ {\rm H}_2/^{13}{\rm CO} \right]$ abundance ratio has to be assumed. In the far outer Galaxy it is expected to
be larger than $5 \times 10^5$ quoted by Dickman (1978) for local dark clouds. Adopting the behaviour of the abundance ratio of
$\left[^{12}{\rm CO}/^{13}{\rm CO}\right]$ {\it versus} the Galactocentric distance suggested by \citet{mil05}, and
assuming a $\left[{\rm H}_2/^{12}{\rm CO}\right]$ abundance ratio of $1.1 \times 10^4$ \citep{fre82}, 
a ratio of $\left[{\rm H}_2/^{13}{\rm CO}\right] = 1.1 \times 10^6$ at $R = 13.6$~kpc is obtained, resulting in a value of $M_\mathrm{LTE}=635~M_{\odot}$ for the mass derived with this method.

Our Hi-GAL-based mass estimate for the clump~E is $M_H=719~M_{\odot}$ (see Table~\ref{hitab}), similar to 
the aforementioned CO-derived masses. Also, as these three mass estimates have the same distance dependence, 
the agreement among them in not affected by the uncertainty in this parameter. Instead, the choice 
of the $X_{CO}$ and $\left[{\rm H}_2/^{13}{\rm CO}\right]$ implies crucial assumptions: the comparison between the CO- and
Hi-GAL-derived masses should be useful to better calibrate these parameters, which in this case, would require relatively
small adjustments. However the $k_\mathrm{ref}$ value adopted in Equation~\ref{mthick} is typically used for the inner Galaxy, 
but the gas-to-dust ratio is expected to increase with decreasing metallicity (conditions that are to be found at large 
Galactocentric radii), leading to large uncertainties on the final gas masses \citep[see][and references therein]{mok07}.
Thus, adopting a larger gas-to-dust ratio in this case would imply, accordingly, a rescaling of $X_{CO}$ and
$\left[{\rm H}_2/^{13}{\rm CO}\right]$.

Once a good agreement between mass estimates obtained from the continuum and line maps has been ascertained, we can discuss the relationship between mass and size, using the quantities in Table~\ref{hitab}. At a distance of 5.5~kpc, the angular extent of the clump E corresponds to a physical diameter of $\simeq 0.4$~pc. This implies a surface density $\Sigma = 1.1$~g~cm$^{-2}$, a value that exceeds the theoretical threshold calculated by \citet{kru08} for a star forming cloud to be able to form massive ($M > 10 M_{\odot}$) stars. This is an interesting finding, testifying the presence of conditions for high mass star formation also in the outer Galaxy, and in relatively isolated regions far from giant star forming clouds. For the clump W, the surface density is found to be even larger, $\Sigma = 1.7$~g~cm$^{-2}$.

The relation between the mass of the envelope (i. e. the mass we derive from Herschel's data) and the bolometric luminosity can be used to diagnose the evolutionary stage of a far-infrared source \citep[e.g.,][]{mol08,ma13,eli13}. The $L_\mathrm{bol}/M$ ratio, in particular, is a distance-independent quantity which is expected to rapidly increase during the accretion phase of star formation.

Here the bolometric luminosity has been obtained as the trapezium-like integral of the observed SED at $\lambda < 160$~$\mu$m (including the fluxes at the WISE bands at 12 and 22 $\mu$m), added to the integral of the best-fitting modified black body at $\lambda \geq 160$~$\mu$m, namely the grey-shaded areas represented in Figure~\ref{seds}. The values obtained for the clumps E and W are reported in Table~\ref{hitab}.
The $L_\mathrm{bol}/M$ ratio amounts to 4.4~$L_{\odot}/M_{\odot}$ for clump E, and to $\sim 1.5 \, L_{\odot}/M_{\odot}$ for clump W. 
These values, corresponding to the formation of a young cluster, cannot be directly compared with models elaborated to describe the formation of a single young stellar object \citep{mol08}. 
However, the direct comparison between clumps E and W shows that the former is likely to be in a more evolved stage. 
Also, the ratio between the bolometric luminosity and its submillimeter portion $L_{\mathrm{submm}}$ derived from fluxes longward of $350~\mu$m, has been used as a further evolutionary indicator, expected to be larger at more evolved star formation stages \citep[e.g.][]{and93}. The values of this ratio, for the E and the W clumps, is found to be 67 and 28, respectively, again confirming the previous indication.
This view is further supported by the fact that: ($i$) clump E has a smaller [70-160] color index than clump W; ($ii$) clump E appears to be slightly warmer than clump W (and the temperature estimate does not depend on the 70 $\mu$m flux); and ($iii$) clump E is less dense than clump W and becoming optically thin at shorter wavelengths (which could indicate that a larger fraction of the gas and dust envelope has already been transferred onto the forming stars, or dissipated by proto-stellar activity).

In Figure~\ref{hls} it can be noticed that the NW clump (cf. Figure~\ref{WISE}) is also partially covered by our $JHK_S$ maps (the red sources at the top right corner of Figure~\ref{JHK} are spatially associated with it) and fully observed by Herschel (clearly detected at all bands, with no duplicities), so that we can derive the physical properties of its dust and gas envelope. A comparison with CO observations, however, is not possible since this source lies outside the area mapped at OSO. The SED of NW is shown in Figure~\ref{seds} and the results of the fit procedure are reported in Table~\ref{hitab}, respectively. This clump appears remarkably fainter and less dense ($\Sigma = 0.2$~g~cm$^{-2}$) than the others. At the same time, it could be going through an earlier evolutionary stage, as testified by its low temperature ($T=12.4$~K) and luminosity/mass ratio ($L_\mathrm{bol}/M=0.3~L_{\odot}/M_{\odot}$), and  by $L_{\mathrm{bol}}/L_{\mathrm{submm}}=10$.

\vspace{50pt}

\section{Summary and Conclusions}

\begin{itemize}
      \item Infrared ($JHK_S$ and {\it Spitzer}) images of the region towards IRAS~04186+5143 reveal a concentration of stars compatible with the presence of a young stellar cluster. 
      \item The cluster is embedded in a molecular cloud core detected through CO, CS, and $NH_3$ line emission. Our CO map reveals the existence of sub-clustering corroborated by the spatial distribution of the young stars.
      \item At 5.5 kpc (heliocentric distance) in the outer Galaxy, and at a galactocentric distance of 13.6~kpc, this population of YSOs may be composed by low and intermediate-mass stars with a large fraction of Class~I sources, a clear sign of a young star formation region.
      \item {\it Herschel} data clearly identify dust clumps coinciding with the positions of the sub-clusters. The {\it Herschel}-derived masses of the main clumps are 719 and 435~$M_{\odot}$, consistent with CO -derived estimates. 
      \item The $L/M$ ratio of the clumps could indicate that the larger (E) clump, hosting a larger fraction of the YSOs seen in the near-infrared images, is in a more evolved stage of the star formation process, having converted more gas into stars than the smaller (W) clump.      
      \item IRAS~04186+5143 is a young stellar cluster forming in the outer Galaxy, and not an external galaxy as identified in the 2MASS extended source catalog and indicated in the SIMBAD data base.
      \item A table is provided giving the photometry of all NOTCam sources detected at least in one near-IR band.
      \item An additional table is provided cross-correlating {\it Spitzer} and NOTCam sources. It contains the mid-infrared photometry of all Spitzer sources with NOTCam $J$, $H$, or $K_S$ counterparts.
   \end{itemize}

\section*{Acknowledgements}
JY acknowledges support from FCT (Portugal) (SFRH/BSAB/1423/2014 and UID/FIS/04434/2013).
JMT acknowledges support from MICINN (Spain) AYA2011-30228-C03 grant (co-funded with FEDER funds).
\textit{Herschel} Hi-GAL data processing, maps production and
source catalogue generation have been possible thanks to Contracts I/038/080/0
and I/029/12/0 from ASI, Agenzia Spaziale Italiana. This work is part of the
VIALACTEA Project, a Collaborative Project under Framework Programme 7
of the European Union, funded under Contract 607380 that is hereby acknowledged.
Based on observations made with the Nordic Optical Telescope, operated on the island of La Palma jointly by Denmark, Finland, Iceland, Norway, and Sweden, in the Spanish Observatorio del Roque de los Muchachos of the Instituto de Astrofisica de Canarias.
This research made use of the NASA/ IPAC Infrared Science Archive, which is operated by the Jet Propulsion Laboratory, California Institute of Technology, under contract with the National Aeronautics and Space Administration.
This research made use of the SIMBAD database, operated at CDS, Strasbourg, France, as well as SAOImage DS9, developed by the Smithsonian Astrophysical Observatory.
This publication made use of data products from the Wide-field Infrared Survey Explorer, which is a joint project of the University of California, Los Angeles, and the Jet Propulsion Laboratory/California Institute of Technology, funded by the National Aeronautics and Space Administration.
\textit{Herschel} is an ESA space observatory with science instruments provided
by European-led Principal Investigator consortia and with important participation
from NASA. PACS has been developed by a consortium of institutes led by
MPE (Germany) and including UVIE (Austria); KUL, CSL, IMEC (Belgium);
CEA, OAMP (France); MPIA (Germany); IAPS, OAP/OAT, OAA/CAISMI,
LENS, SISSA (Italy); IAC (Spain). This development has been supported by
the funding agencies BMVIT (Austria), ESA-PRODEX (Belgium), CEA/CNES
(France), DLR (Germany), ASI (Italy), and CICYT/MCYT (Spain). SPIRE has
been developed by a consortium of institutes led by Cardi Univ. (UK) and including
Univ. Lethbridge (Canada); NAOC (China); CEA, LAM (France); IAPS,
Univ. Padua (Italy); IAC (Spain); Stockholm Observatory (Sweden); Imperial
College London, RAL, UCL-MSSL, UKATC, Univ. Sussex (UK); Caltech, JPL,
NHSC, Univ. Colorado (USA). This development has been supported by national
funding agencies: CSA (Canada); NAOC (China); CEA, CNES, CNRS (France);
ASI (Italy); MCINN (Spain); Stockholm Observatory (Sweden); STFC (UK);
and NASA (USA).

\bibliography{biblio_higal}
\bibliographystyle{mn2e}

\label{lastpage}

\end{document}